\documentclass[fleqn,usenatbib]{mnras}

\usepackage{newtxtext,newtxmath}

\usepackage[T1]{fontenc}
\usepackage{ae,aecompl}

\usepackage{graphicx}	
\usepackage{pifont}     
\usepackage{subfiles}   
\usepackage{blindtext}  

\title[HII region trapping in Pop III star formation]{Trapping of HII regions in Population III star formation}

\author[Jaura et al.]{Ondrej Jaura$^{1}$,
Simon C. O. Glover$^{1}$,
Katharina M. J. Wollenberg$^{1}$,
\newauthor
Ralf S. Klessen$^{1,2}$,
Sam Geen$^{1,3}$,
Lionel Haemmerl\'e$^{1,4}$
\\
$^{1}$Universit\"{a}t Heidelberg, Zentrum f\"{u}r Astronomie, Institut f\"{u}r Theoretische Astrophysik, Albert-Ueberle-Str. 2, 69120 Heidelberg, Deutschland \\
$^{2}$Universit\"at Heidelberg, Interdisziplin\"ares Zentrum f\"ur Wissenschaftliches Rechnen, Im Neuenheimer Feld 205, 69120 Heidelberg, Germany \\
$^{3}$Anton Pannekoek Institute for Astronomy, Universiteit van Amsterdam, Science Park 904, 1098 XH Amsterdam, The Netherlands\\
$^{4}$D\'epartement d'Astronomie, Universit\'e de Gen\'eve, chemin des Maillettes 51, CH-1290 Versoix, Switzerland
}

\date{Accepted XXX. Received YYY; in original form ZZZ}

\pubyear{2022}

\begin{document}
\label{firstpage}
\pagerange{\pageref{firstpage}--\pageref{lastpage}}
\maketitle

\begin{abstract}
Radiative feedback from massive Population III (Pop.\ III) stars in the form of ionising and photodissociating photons is widely believed to play a central role in shutting off accretion onto these stars. Understanding whether and how this occurs is vital for predicting the final masses reached by these stars and the form of the  Pop.~III stellar initial mass function. To help us better understand the impact of UV radiation from massive Pop.~III stars on the gas surrounding them, we carry out high resolution simulations of the formation and early evolution of these stars, using the AREPO moving-mesh code coupled with the innovative radiative transfer module SPRAI. Contrary to most previous results, we find that the ionising radiation from these stars is trapped in the dense accretion disk surrounding them. Consequently, the inclusion of radiative feedback has no significant impact on either the number or the total mass of protostars formed during the 20~kyr period that we simulate. We show that the reason that we obtain qualitatively different results from previous studies of Pop.~III stellar feedback lies in how the radiation is injected into the simulation. HII~region trapping only occurs if the photons are injected on scales smaller than the local scale height of the accretion disk, a criterion not fulfilled in previous 3D simulations of this process. Finally, we speculate as to whether outflows driven by the magnetic field or by Lyman-$\alpha$ radiation pressure may be able to clear enough gas away from the star to allow the HII~region to escape from the disk. 
\end{abstract}




\section{Introduction}

The appearance of the first stars ended the so-called ``dark ages'' of our Universe \citep{loeb_how_2010}. They played a key role in cosmic metal enrichment and reionisation, thereby shaping the galaxies and their internal properties as we see them today. In order to study the impact of the first generations of stars, the so-called Population III (or Pop III for short), on subsequent cosmic evolution we need to know when and where they form, and how many to expect. We also need to better understand the physical processes that control their initial mass function (IMF) and multiplicity. These are the primary parameters that determine the spectral energy distribution of the stellar radiation field, the amount and composition of heavy elements produced, and the energy and momentum carried away by winds and eventually by supernova explosions. 

Population III stars form by gravitational collapse of truly metal-free primordial gas that is accumulated in the potential wells of dark matter halos. Population III star formation is expected to begin at redshifts $z \gtrsim 30$ and to reach a peak at $z \sim 15$--20 \citep[see e.g.][]{hummel_source_2012,magg_new_2016}. Although the overall cosmic star formation rate continues to increase at lower redshifts \citep{madau_super-critical_2014}, the rate at which metal-free stars form declines at later times as gas that is not enriched by supernova ejecta becomes increasingly rare  \citep{bromm_first_2004,yoshida_formation_2012,bromm_formation_2013, glover_first_2013, klessen_formation_2018}. Early studies of Pop III star formation predicted that only one extremely massive star with $M > 100 \: {\rm M_{\odot}}$ should form in each dark matter halo \citep{omukai_primordial_2001,abel_formation_2002,bromm_formation_2002,oshea_population_2007}. With the ever-increasing capabilities of modern supercomputers, however, this situation has changed, and more recent investigations and numerical simulations lead to the conclusion that fragmentation is a widespread phenomenon in first star formation \citep{clark_gravitational_2011, greif_formation_2012}. We now believe that most Pop III stars form as members of multiple stellar systems with a wide range of separations and mass ratios \citep{turk_formation_2009, clark_formation_2011, greif_delay_2011, smith_effects_2011, stacy_constraining_2013}. This raises the question of whether these fragments survive or merge together. As yet, there is no convincing answer to this question because all existing analytic or numerical models either deal with restricted geometry, only include a subset of the relevant physical processes, or only cover the initial phase of the overall evolution. Studies that do include radiative feedback \citep{omukai2002,mckee_formation_2008,hirano_one_2014, hirano_primordial_2015, hosokawa_formation_2016,stacy_building_2016,sugimura_birth_2020}, magnetic fields \citep{machida_first_2006, machida_magnetohydrodynamics_2008, schleicher_influence_2009, sur_generation_2010, sur_magnetic_2012, turk_magnetic_2012, schober_small-scale_2012, schober_magnetic_2012, bovino_turbulent_2013,sharda2020,Sharda2021},
dark matter annihilation \citep{smith_variable_2012, stacy_mutual_2014}, as well as the primordial streaming velocities \citep{tseliakhovich_relative_2010, greif_delay_2011, maio_impact_2011, stacy_effect_2011, schauer_influence_2019} add to this complexity. 

In this paper, we study the impact of radiative feedback on the formation of the first stars, with specific emphasis on resolving the immediate vicinity of the assembling stars with very high resolution. Since the protostellar Kelvin-Helmholtz contraction time decreases rapidly with increasing stellar mass, massive stars enter the hydrogen burning main sequence while still accreting \citep{zinnecker_toward_2007,maeder_rotating_2012}. The resulting stellar parameters strongly depend on the details of the mass growth history, stressing again the importance of properly resolving the accretion flow onto the protostar in numerical simulations. In the low-mass halos investigated here, with typical accretion rates below  $\dot{M} \approx 10^{-3}\,$M$_{\odot} \, {\rm yr^{-1}}$, the resulting Pop III stars are compact and very hot at their surface \citep{hosokawa_evolution_2009, hosokawa_evolution_2010, hosokawa_protostellar_2012}. They emit large numbers of Lyman-Werner and ionising photons \citep{schaerer_properties_2002} that can significantly influence their birth environment through the mechanisms that we discuss in Section~\ref{sec:sprai}. Consequently, the question of how long the resulting HII regions (volumes of gas containing ionised hydrogen) remain small and compact, and when or whether they break out of the parental halo, is crucial for our understanding of how they might affect stellar birth in neighboring halos. Many aspects of this problem have been addressed, for example by \citet{kitayama_structure_2004}, \citet{whalen_radiation_2004}, \citet{alvarez_h_2006}, \citet{abel_h_2007}, \citet{yoshida_early_2007}, \citet{greif_first_2008}, \citet{wise_birth_2012-1}, \citet{wise_birth_2012}, or \citet{jeon_radiative_2014}. Here we focus on the impact of radiative feedback on the immediate birth environment of the star. 

When studying this problem in primordial star formation, we can seek guidance from models of stellar birth at the present day. Radiation hydrodynamic simulations in 2D and 3D \citep{yorke_formation_2002,krumholz_formation_2009, kuiper_circumventing_2010, kuiper_three-dimensional_2011, peters_understanding_2010, peters_interplay_2011, commerccon_collapse_2011, rosen_unstable_2016} demonstrate that once a protostellar accretion disk has formed, it quickly becomes gravitationally unstable and so material in the disk midplane flows inwards along dense filaments, whereas  radiation escapes through optically thin channels above the disk. Even ionised material can be accreted, if the accretion flow is strong enough \citep{keto_formation_2007}. Radiative feedback is thought not to be able to shut off the accretion flow onto massive stars. Instead it is the dynamical evolution of the disk material that controls the mass growth of individual protostars. Accretion onto the central object is shut off by the fragmentation of the disk and the formation of lower-mass companions which intercept inward-moving material \cite[see, e.g.][]{girichidis_importance_2012}. This requires 3D simulations, since the fragmentation process is not properly captured in two dimensions.

Due to the lack of metals and dust, the accretion disks around Pop III stars can cool less efficiently and are much hotter than present-day protostellar accretion disks \citep{tan_formation_2004, glover_formation_2005}. Similarly, the stellar radiation field couples less efficiently to the  surroundings because the opacities are smaller \citep{glover_chemistry_2011}. It is thus not clear how well the above results can be transferred to the primordial case. Radiation hydrodynamic simulations in 2D, for example, have been presented by \citet{hosokawa_protostellar_2011}, \citet{hirano_one_2014}, and \citet{hirano_primordial_2015}. They find that radiative feedback can indeed stop stellar mass growth and blow away the accretion disk, resulting in large HII regions that break out of the parental halo. They also report final stellar masses in the range from a few $10\,$M$_\odot$ up to about $1000\,$M$_\odot$. However, these calculations cannot capture disk fragmentation and the formation of multiple stellar systems. Three-dimensional calculations have been reported by \citet{stacy_first_2012}, \citet{susa_mass_2013}, \citet{susa_mass_2014}, \citet{hosokawa_formation_2016}, \citet{stacy_building_2016}, and \citet{sugimura_birth_2020}. These studies find widespread fragmentation, again with a wide range of stellar masses down to $\sim 1\,$M$_\odot$. Again, they find that HII regions eventually break out and grow extremely large. However, a significant limitation of all of these 3D studies is the physical resolution achieved in the vicinity of the massive star. This ranges from $\sim 10$--50~AU, which is significantly larger than the scale height of the protostellar accretion disk close to the star. Therefore, none of these existing calculations properly represents the interaction between the ionising radiation and the inner disk. 

Our goal here is to improve on these earlier studies by simulating the formation of massive Pop III stars with sufficient resolution to follow the interaction between the ionising radiation and the gas in the inner regions of the disk. We show that this leads to a qualitative difference in the behaviour of the HII region, with important implications for the outcome of the Pop III star formation process.

Our paper is structured as follows. Section~\ref{sec:numerical-methods} describes the numerical methods used in our simulation. In Section~\ref{sec:initial-conditions}, we introduce our set of initial conditions and simulation settings. Results of the simulations are given in Section~\ref{sec:results}. After that, in Section~\ref{sec:discussion} we discuss our results, along with some caveats of our model and compare the results of our simulations with others from the recent literature. We conclude in Section~\ref{sec:conclusions}.

\section{Numerical Methods}
\label{sec:numerical-methods}

For our simulations we use a version of the cosmological hydrodynamical code \textsc{Arepo} \citep{springel_e_2010} that incorporates several modules that we have developed for modelling Pop.~III star formation. These include a detailed model of the chemistry and thermal physics of primordial gas, collisionless sink particles, which we use to represent individual Pop.~III stars, and the {\sc sprai} radiative transfer module (\textsc{sprai-i}: \citealt{jaura_sprai:_2018}; \textsc{sprai-ii}: \citealt{jaura_sprai-ii_2020}). The latter is a novel treatment of the effects of ionising and photodissociating radiation based on the \textsc{SimpleX} algorithm \citep{kruip_mathematical_2010,paardekooper_simplex2:_2010}. In this section, we describe each of these modules in more detail. 

\subsection{Chemistry and thermal physics}
\label{sec:popiii-chemistry}

To model the chemistry of primordial gas, we use a similar chemical network to the one used by \citet{wollenberg_formation_2019}. This network is based on the one described in detail in \citet{clark_gravitational_2011}, but includes several updates to the chemical rate coefficients, as outlined in \citet{schauer_formation_2017}. The only significant difference between the chemical network used in our current study and the one adopted in these previous studies is the inclusion of the effects of photodissociating and photoionising radiation from massive Pop.\ III stars. We include four main processes: Lyman-Werner photodissociation of H$_{2}$, and photoionisation of H, He and H$_{2}$. The first three of these correspond to the reactions
\begin{eqnarray}
{\rm H_{2} + \gamma} & \rightarrow & {\rm H + H}, \\
{\rm H + \gamma} & \rightarrow & {\rm H^{+} + e^{-}}, \\
{\rm He + \gamma} & \rightarrow & {\rm He^{+} + e^{-}}.
\end{eqnarray}
The photoionisation of H$_{2}$ in principle should be represented by the reaction
\begin{eqnarray}
{\rm H_{2} + \gamma} & \rightarrow & {\rm H_{2}^{+} + e^{-}},
\end{eqnarray}
but in the dense ionised gas that we are concerned with in this study, we assume that all of the resulting H$_{2}^{+}$ will rapidly be destroyed by dissociative recombination, resulting in the production of two hydrogen atoms. We therefore include this process using the pseudo-reaction
\begin{eqnarray}
{\rm H_{2} + \gamma} & \rightarrow & {\rm H + H}.
\end{eqnarray}
The reaction rates for these four processes are computed using the {\sc sprai} radiation transfer module, described in Section~\ref{sec:sprai} below. In principle, we should also account for the photodetachment of H$^{-}$ and photodissociation of H$_{2}^{+}$ by photons from the massive Pop.\ III stars. However, in practice, we do not expect these reactions to be important in the conditions simulated here, since we are primarily interested in the behaviour of the dense gas close to the stars, and in this dense gas, three-body formation of H$_{2}$ dominates over formation via the H$^{-}$ or H$_{2}^{+}$ pathways \citep{palla_primordial_1983}.

As in \citet{wollenberg_formation_2019}, we also make the simplification that in gas denser than $n = 10^{8} \: {\rm cm^{-3}}$, the HD/H$_{2}$ ratio and atomic D/H ratio are both simply equal to the cosmological D to H ratio, $x_{\rm D, tot} = 2.6 \times 10^{-5}$, enabling us to neglect the portions of the chemical network involving deuterium. At high densities, the reactions responsible for converting H$_{2}$ to HD (or vice versa) and for transferring charge from H$^{+}$ to D and from D$^{+}$ to H are extremely rapid, and modelling them accurately in the chemical module is therefore computationally costly, and yet results in HD/H$_{2}$, D/H etc.\ ratios that are very close to the cosmological D to H ratio. Making this simplification therefore allows us to substantially speed up our simulations while resulting in very little difference in the behaviour of the gas. As a consequence of this simplification, we do not explicitly treat the photoionisation of D or photodissociation of HD in our model, but instead assume that both processes occur at the same rate as for H and H$_{2}$, respectively. Note that our simplification here only concerns the HD {\em chemistry}: we continue to include its contribution to the radiative cooling at $n > 10^{8} \: {\rm cm^{-3}}$ and merely use a much simpler scheme for tracking its abundance than the full non-equilibrium treatment we adopt at lower densities.

As well as the chemical evolution, we also solve simultaneously for the thermal evolution of the gas due to the effects of radiative and chemical heating and cooling. To model radiative cooling from H$_{2}$ rotational and vibrational line emission, we use the detailed H$_{2}$ cooling function presented in \citet{glover_uncertainties_2008}, updated as described in \citet{glover_simulating_2015}. The effects of H$_2$ line opacity in dense gas are accounted for using the modified Sobolev approximation introduced by \citet{clark_formation_2011}. Cooling due to collision-induced emission (CIE) from H$_2$ is accounted for using an optically thin rate from \citet{ripamonti_fragmentation_2004} and the opacity correction described in \citet{clark_formation_2011}. For HD cooling, we use the temperature and density-dependent rate given in \citet{lipovka_cooling_2005}. We also account for radiative cooling due to electronic excitation of H, He and He$^{+}$, the recombination of H$^{+}$ and He$^{+}$, Compton cooling and bremsstrahlung using the rates given in \citet{glover_star_2007}. Radiative heating due to the photoionisation of H, H$_{2}$ and He and the photodissociation of H$_{2}$ is computed using {\sc sprai} (see later). Chemical heating due to H$_{2}$ formation and chemical cooling due to the collisional dissociation of H$_{2}$ and the collisional ionisation of H and He are modelled as in \citet{clark_formation_2011}.

Finally, we account for the fact that in gas with a significant molecular fraction, the adiabatic index $\gamma$ varies as a function of the temperature and H$_2$ fraction. We use the HLLD\footnote{Note that HLLD is a magnetohydrodynamic solver, but in our simulation the magnetic field is switched off.} Riemann solver built into {\sc Arepo}  \citep{pakmor_magnetohydrodynamics_2011}, which supports the use of a variable $\gamma$, and compute the variation of $\gamma$ with temperature and chemical composition using the same approach as in \citet{boley_internal_2007}.

\subsection{Sink particles}
\label{sec:popiii-sinks}

Ideally, when modelling the formation of Pop.\ III stars, we would like to be able to follow the gravitational collapse of the gas down to the scale of the individual stars themselves. Unfortunately, the computational demands of doing so are extremely high. For example, \citet{greif_formation_2012} follow the collapse of primordial gas down to scales of less than a solar radius, corresponding to protostellar densities. However, as a consequence, they are only able to follow the evolution of the system for $\sim 10$~yr. This is orders of magnitude shorter than the time required to form even a single massive star, making this approach impractical for our purposes. Instead, it is necessary to use an approach in which the collapse of the gas is not followed down to such small length scales. 

There are two main ways in which this can be accomplished. One possibility is to force the gravitational collapse of the gas to stop earlier than it would do in reality, by modifying either the equation of state or the cooling function \citep[see e.g.][]{vorobyov_burst_2013,machida_accretion_2015,hosokawa_formation_2016,hirano_formation_2017,susa_merge_2019}. Alternatively, collapsing regions can be removed from the simulation entirely and replaced with collisionless sink particles or sink cells \citep[e.g.][]{clark_formation_2011,susa_mass_2014,stacy_building_2016}.

In our simulations, we have chosen to use the latter approach. Our sink particle model is described in detail elsewhere \citep{wollenberg_formation_2019} and so here we give only a few basic details. {\sc arepo} grid cells become eligible for conversion into sink particles if they have densities exceeding a threshold density $n_{\rm th}$ and are also situated at a local minimum of the gravitational potential. In addition, the gas within a sphere of radius $r_{\rm sink}$ (the accretion radius) around the candidate cell must be gravitationally bound and collapsing. Finally, sink formation is suppressed in cells that are already closer than $r_{\rm sink}$ to an existing sink particle. 

Once created, sink particles can accrete gas from any cells that are located at a distance $r \leq r_{\rm sink}$ from the sink. The gas must be gravitationally bound to the sink and the cell must have a density $n > n_{\rm th}$. If the cell is located within a distance of $r_{\rm sink}$ from more than one sink particle, then the gas in the cell is only eligible for accretion by the sink to which it is most tightly bound. If these conditions are satisfied, then the sink accretes sufficient gas from the cell to reduce the cell density to $n_{\rm th}$ (or 90\% of the total cell gas mass if this is smaller). In the simulations presented in this paper, we adopt a density threshold $n_{\rm th} = 7.248\times10^{13} \, {\rm cm}^{-3}$. We carry out simulations with three different values of the sink accretion radius, 2, 10 and 30~AU, as summarized in Table~\ref{tab:init-settings} and discussed in more detail in Section \ref{sec:initial-conditions}. As in \citet{wollenberg_formation_2019}, we do not account for mergers between sink particles in our current study. The impact of including mergers is difficult to assess, since it depends sensitively on the criterion used to determine whether two sinks merge, but would likely lead to the formation of somewhat more massive stars than those that form in the simulations presented here.

As our sinks represent individual protostars (or, later, main sequence stars), it is also necessary to account for the energy released as gas accretes onto their surfaces. The contribution that this makes to the flux of ionising and photodissociating photons is accounted for using the method described in Section~\ref{sec:emission-from-popiii} below. However, in addition, it is also necessary to account for the heating of the gas by the accreting protostars prior to them reaching the main sequence. This accretion luminosity is typically characterised by a low radiation temperature and hence does not contribute to the aforementioned fluxes, but nevertheless is important for regulating the fragmentation of the gas \citep{smith_effects_2011}. We treat the effects of this accretion luminosity heating using the method described in detail in \citet{wollenberg_formation_2019}, which itself is a slightly modified version of an approach first used in \citet{smith_effects_2011}. Briefly, we use a simple model of the early evolution of the protostars to solve for their bolometric luminosity as a function of their mass and their current accretion rate. Since the effective temperatures of the protostars before they join the main sequence are $T_{\rm eff} \ll 10^{4}$~K, most of the photons emitted during this phase have energies significantly below the hydrogen photoionisation threshold. For this range of photon energies, the continuum opacity of primordial gas is very small at the densities encountered in our study (\citealt{mayer_rosseland_2005}; see also the detailed calculations of continuum optical depths as a function of density in \citealt{clark_formation_2011} and \citealt{HiranoYoshida2013}). It is therefore a good approximation to take the gas surrounding the protostars to be optically thin to their continuum emission. We can therefore write the flux of radiation at a distance $R$ from a protostar with accretion luminosity $L$ simply as $F = L / (4 \pi R^{2})$. The heating rate of the gas due to this flux then follows from
\begin{equation}
\Gamma_{\rm acc} = \rho \kappa_{\rm P}(\rho, T) F,
\end{equation}
where $\rho$ is the gas density and $\kappa_{\rm P}$ is the Planck mean opacity of the gas at its current density and temperature, which we interpolate from the values tabulated by \citet{mayer_rosseland_2005}. Stars with masses $M > 10 \: {\rm M_{\odot}}$ that are contracting to the main sequence have effective temperatures $T_{\rm eff} > 10^{4}$~K and so emit significant numbers of ionising and photodissociating photons, which are treated as described in the next section. However, even for these stars, the assumption of optically thin gas remains valid for photons with energies below 11.2~eV at the densities we resolve in our current study \citep{mayer_rosseland_2005}. Note that because of our sink accretion procedure, gas within the accretion radius generally remains at a density close to $n_{\rm th}$, whereas in reality some fraction of the gas in this region would likely be denser. It is therefore possible that some of the gas very close to the star could actually be optically thick in the continuum, although verifying this would require simulations with a higher resolution than we can currently afford. However, as this gas would simply re-radiate the majority of the radiation it receives from the protostar, its presence would not have a major impact on the value of $\Gamma_{\rm acc}$ at distances $r > r_{\rm sink}$ and hence would not significantly affect the outcome of our calculations.

\subsection{Modelling ionising and photodissociating radiation with {\sc sprai}}
\label{sec:sprai}
Initially, the accreting protostars formed in our simulations have large radii and low effective temperatures \citep{stahler_primordial_1986,omukai_formation_2001,omukai_formation_2003} and primarily affect their surroundings via the accretion luminosity heating discussed in the previous section. However, once their accretion timescale exceeds the Kelvin-Helmholtz timescale, the stars contract and increase their effective temperature, until they join the zero-age main sequence (see Figure~\ref{fig:spectrum-temp-rad}). As a consequence, the star becomes a source of both ionising and photodissociating photons, provided its mass is large enough. 

To manage the transport of ionising and photodissociating radiation through the Voronoi mesh cells in our simulations and to compute the resulting photochemical and heating rates, we use the {\sc sprai} radiation transfer module, described in \citet{jaura_sprai:_2018} and \citet{jaura_sprai-ii_2020}. {\sc sprai} is a ray-tracing method based on the {\sc SimpleX} algorithm of
\citet{kruip_mathematical_2010} and \citet{paardekooper_simplex2:_2010}, which is a variant of the short characteristics approach. At the beginning of each full hydrodynamical timestep\footnote{A full hydrodynamical timestep in {\sc arepo} is one on which all of the mesh cells are synchronized.}, \textsc{sprai} calculates the number of photons emitted by each source in each of the tracked energy bins, using the procedure described in Section~\ref{sec:emission-from-popiii} below. These photons are distributed equally amongst a set of directional bins associated with the gas cell in which the source resides. The photons are then moved step by step on the mesh along paths that approximately follow the directions defined by the directional bins. The effects of attenuation are computed for each cell that the photons pass through and the photons are propagated until they are completely attenuated. Finally, the number of photons in each energy bin absorbed in each cell during the timestep is used to calculate the corresponding photoionisation, photodissociation and photoheating rates, which are passed on to the chemistry module. The change in the momentum of the gas in the cell owing to the absorption of the photons -- i.e.\ the effect of the radiation pressure -- is also accounted for at this stage.

A strength of {\sc sprai} in comparison to more conventional long characteristics methods \citep[e.g.][]{wise_enzomoray_2011} is that every cell can potentially act as a source cell and hence the computational cost of the method is determined by the number of cells that are ionised and not by the number of ionising sources. This makes it a good choice for situations containing multiple sources of radiation, as in the simulations described later in this paper. It should be noted that this flexibility comes at a cost: short characteristics methods are generally more diffusive than long characteristics methods, and {\sc sprai} is no exception. Nevertheless, it still proves capable of modelling effects such as shadowing with minimal leakage of radiation into the shadowed region \citep[see e.g.\ test 4 in][]{jaura_sprai:_2018}. A more comprehensive description of {\sc sprai} and the results of a series of tests of the method can be found in \citet{jaura_sprai:_2018} and \citet{jaura_sprai-ii_2020}.

\subsubsection{Emission from accreting Population III stars}
\label{sec:emission-from-popiii}

\begin{figure}
  \includegraphics[width=\linewidth]{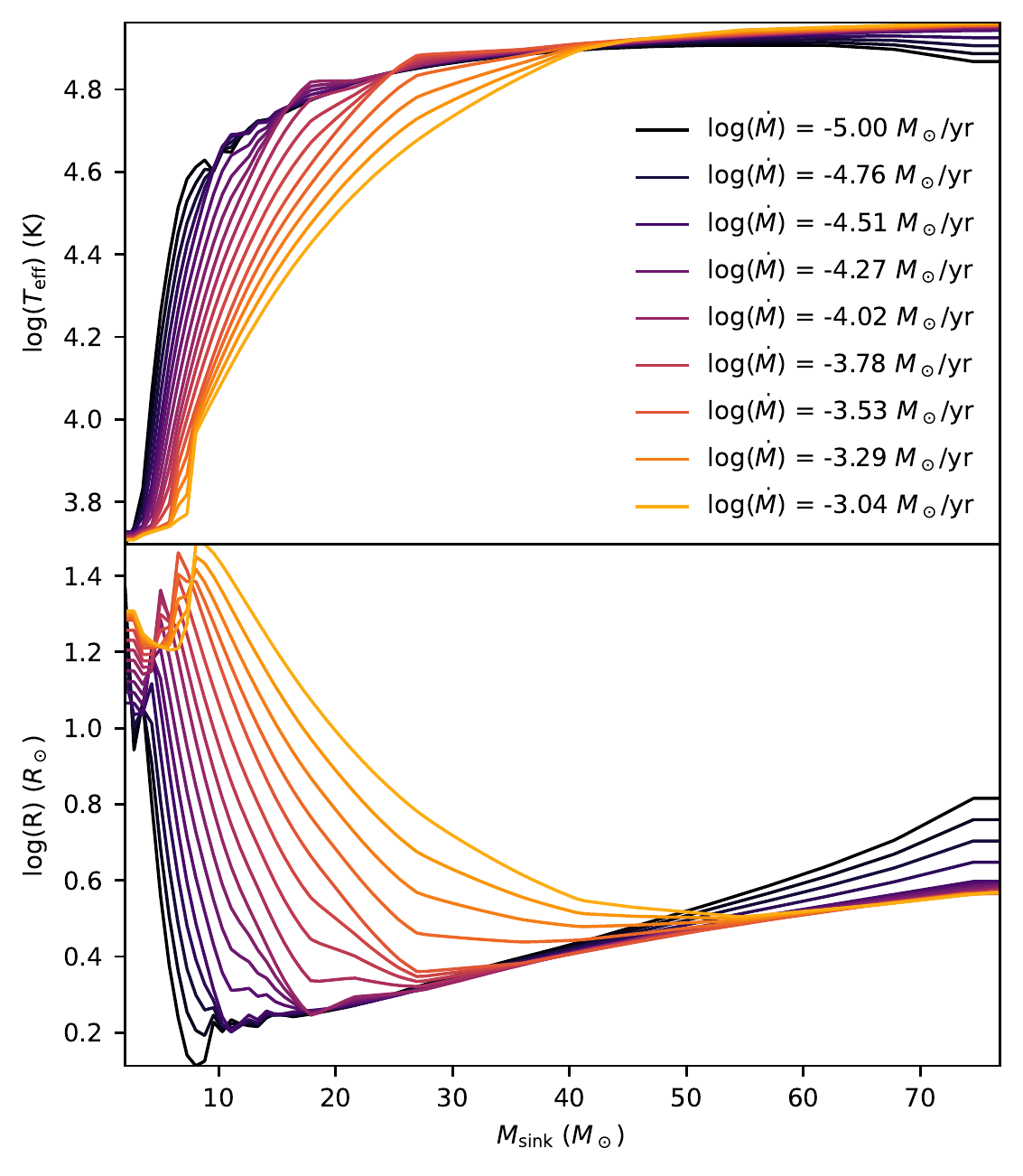}
  \caption{Dependence of the stellar effective temperature (top) and the stellar radius (bottom) on the mass of the star and its accretion rate. These values are based on the Pop III stellar models presented in \citet{haemmerle_evolution_2018}.}
  \label{fig:spectrum-temp-rad}
\end{figure}

The sink particles formed in our simulations are taken to represent individual Pop.\ III stars, and therefore each one act as a source of radiation. To model the emission from each sink, we make use of the Pop.\ III stellar models presented in \citet{haemmerle_evolution_2018}. These models use the current mass of the star together with its mass accretion rate in order to compute the stellar radius $R_{*}$ and effective temperature $T_{\rm eff}$ of the accreting stars. The total luminosity then follows trivially from the Stefan-Boltzmann equation:
\begin{equation}
L = 4\pi \sigma_{\rm SB} R_{*}^{2} T_{\rm eff}^{4},
\end{equation}
where $\sigma_{\rm SB}$ is the Stefan-Boltzmann constant. The dependence of $R_{*}$ and $T_{\rm eff}$ on the stellar mass and accretion rate is illustrated in Figure~\ref{fig:spectrum-temp-rad}. Each Pop.\ III star starts its evolution on the Hayashi line, with an effective temperature of around 5000~K and a radius $R_{*} > 10 \, {\rm R}_\odot$. It maintains a low $T_{\rm eff}$ and large $R_*$ until the Kelvin-Helmholtz time becomes shorter than the accretion time $M/\dot M$. At this point, the star contracts efficiently and converges to the zero-age main sequence, with an effective temperature near $10^5$~K for the most massive stars. The transition between these `red' and `blue' regimes occurs at larger masses for larger accretion rates because of a shorter accretion time, i.e.\ larger luminosities are required for efficient contraction. 

In order to determine $R_{*}$ and $T_{\rm eff}$ for a given star, we therefore need to calculate the current accretion rate onto the star. To do this, we divide the change in mass of the sink over the last full hydrodynamical timestep by the size of the timestep.\footnote{In practice, this approach likely over-estimates the variability in $R_{*}$ and $T_{\rm eff}$ at early times compared to an approach where the accretion rate is averaged over a longer period; see e.g.\ the analysis of this effect in \citet{smith_variable_2012}.} Note that in all of our simulations, following the formation of sink particles, we limit the size of this timestep to be no larger than $\Delta t_\mathrm{max} = 1 \: {\rm yr}$. For accretion rates 
outside of the tabulated range ($10^{-5}$--$10^{-3} \: {\rm M_{\odot} \, yr^{-1}}$), we use values of $T_{\rm eff}$ and $R_{*}$ corresponding to either the largest or the smallest value of $\dot{M}$ within the tabulated range, as appropriate.

Given the luminosity and effective temperature of each star, we can then compute the number of photons per unit time that each emits in the various photon energy bins tracked by {\sc sprai}. As discussed in more detail in \citet{jaura_sprai-ii_2020}, we currently track photons in four bins with the following energy ranges:
\begin{equation*}
\begin{array}{ccccc}
 \hspace{.9in}   11.2 & < & E & < & 13.6 \: {\rm eV}  \\ 
 \hspace{.9in}   13.6 & < & E & < & 15.2 \: {\rm eV} \\
 \hspace{.9in}   15.2 & < & E & < & 24.6 \: {\rm eV} \\
         & & E & > & 24.6 \: {\rm eV}
\end{array}
\end{equation*}
For consistency with \citet{jaura_sprai-ii_2020}, we will refer to these as the 11.2+, 13.6+, 15.2+ and 24.6+ bins, respectively. 

\begin{table}
\begin{center}
\begin{tabular}{ccccc}
	Bin (eV) & H$_2^\mathrm{dis}$ & H$^\mathrm{ion}$ & H$_2^\mathrm{ion}$ & He$^\mathrm{ion}$\\ \hline
	11.2 - 13.6 & \ding{51} & - & - & -  \\
	13.6 - 15.2 & \ding{51} & \ding{51} & - & - \\ 
	15.2 - 24.6 & - & \ding{51} & \ding{51} & -  \\
	24.6+ & - & \ding{51} & \ding{51} & \ding{51} \\
\end{tabular}
\caption{List of energy bins and the processes for which they are responsible.}
\label{tab:freq-bins}
\end{center}
\end{table}

\begin{figure}
  \includegraphics[width=\linewidth]{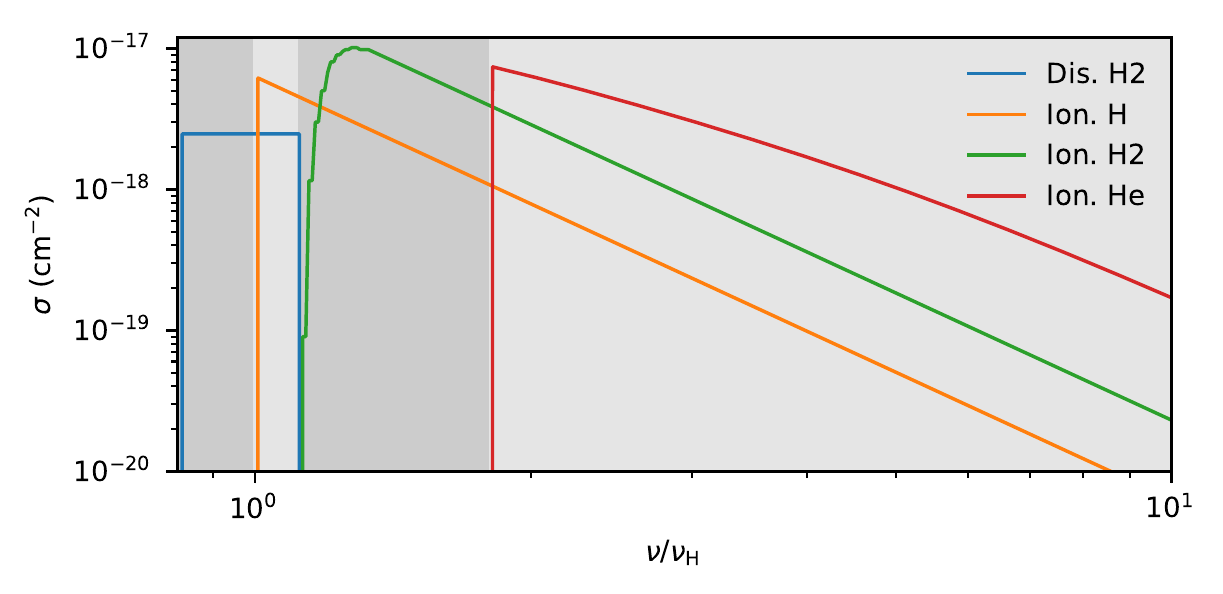}
  \caption{Cross-sections adopted for the different photochemical processes included in our chemical model. The shaded regions correspond to the different energy bins: from left to right, we have the 11.2+, 13.6+, 15.2+ and 24.6+ bins, respectively.}
  \label{fig:freq-cross-sections}
\end{figure}

Our choice of energy bins is motivated by the fact that the four photochemical processes we are interested in tracking -- H$_{2}$ photodissociation and the photoionisation of H, H$_{2}$ and He -- have different energy thresholds. Table~\ref{tab:freq-bins} summarizes which bins correspond to which processes, and Figure~\ref{fig:freq-cross-sections} shows the cross-section adopted for each process. The cross-sections we use for the photoionisation of H, H$_{2}$ and He are taken from \citet{osterbrock_astrophysics_1989}, \citet{liu_nondissociative_2012} and \citet{verner_atomic_1996}, respectively. 

For H$_{2}$ photodissociation, we follow \citet{baczynski_fervent:_2015} and use an effective, frequency-independent cross-section derived by taking the ratio of the photodissociation rate to the Lyman-Werner photon flux in the optically thin limit. We discuss the limitations of this approach in more detail in Section~\ref{methods:lyman-werner} below.

\subsubsection{Treatment of Lyman-Werner radiation}
\label{methods:lyman-werner}
H$_{2}$ photodissociation by photons in the Lyman-Werner bands of molecular hydrogen is a two-stage process driven by line absorption, and is hence inherently more complicated to model than the photoionisation of H, H$_{2}$ or He. Accurately modelling the absorption of radiation in each of the individual Lyman and Werner lines would require an extremely high number of frequency bins and hence is not computationally feasible in the context of our simulations. Instead, we follow \citet{baczynski_fervent:_2015} and adopt a frequency-independent cross-section of $\sigma_{\rm H_{2}, dis} = 2.47 \times 10^{-18} \: {\rm cm^{2}}$ for this process. \citet{baczynski_fervent:_2015} derived this value by taking the ratio of the photodissociation rate computed by \citet{rollig_photon_2007} for the \citet{draine_photoelectric_1978} interstellar radiation field and the photon flux given by \citet{draine_structure_1996} for the same field. Although the \citet{draine_photoelectric_1978} field is unlikely to be a particularly good representation of the ultraviolet radiation field produced by massive Pop.\ III stars, the ratio of the photodissociation rate and photon flux varies only weakly with changes in the spectral shape, and so our adopted value of $\sigma_{\rm H_{2}, dis}$ should still be a reasonable representation of the behaviour we expect in the optically thin regime.

A more significant problem is the fact that we expect $\sigma_{\rm H_{2}, dis}$ to decrease as the H$_{2}$ column density between the gas and the source of radiation increases, owing to the increasing effectiveness of H$_{2}$ self-shielding. This is commonly accounted for through the use of a self-shielding function that is a function of the H$_2$ column density (see e.g. \citealt{draine_structure_1996}, \citealt{wolcott-green_suppression_2011}). Unfortunately, this approach is not viable in {\sc sprai}, as the Lyman-Werner photons entering a cell carry no information with them on their source (or sources), and so it is not clear what H$_2$ column density should be used to compute the self-shielding function. We deal with this problem simply by ignoring the self-shielding correction and assuming that $\sigma_{\rm H_{2}, dis}$ remains unaltered regardless of whether we are in the optically thin or optically thick regimes. Consequently, we under-estimate the reduction of the H$_{2}$ photodissociation rate at moderate H$_{2}$ column densities and over-estimate it at high H$_{2}$ column densities. However, this will have a significant impact on our results only in circumstances where we resolve the H$_{2}$ photodissociation front. 
We can estimate the required spatial resolution by determining the resolution required to ensure that the H$_{2}$ column density within a single cell is less than $10^{21} \: {\rm cm^{-2}}$, since column densities of H$_{2}$ greater than this result in almost complete absorption of all radiation within the Lyman-Werner bands \citep{draine_structure_1996}. This requirement yields a length scale of approximately $\Delta x \sim 1 x_{\rm H_{2}}^{-1} (n / 10^{8} \: {\rm cm^{-3}})^{-1} \: {\rm AU}$. In the dense, fully molecular gas found within the protostellar accretion disk, this length scale is much smaller than the cell size, and so in these regions our simplification should have little impact on our results. On the other hand, in the partially molecular gas above and below the disk, this length scale can become larger than the cell size, and so here our simplification likely leads to us somewhat over-estimating the impact of the Lyman-Werner radiation.

Finally, in addition to the absorption of Lyman-Werner photon by H$_{2}$, we also account for the absorption of photons in the 11.2+ bin by the Lyman series lines of atomic hydrogen. This can become important when the atomic hydrogen column density is very large \citep{wolcott-green_suppression_2011,glover_dynamics_2017,schauer_lyman-werner_2017} and in dense gas acts to prevent the photodissociation front from advancing much beyond the ionisation front \citep{glover_dynamics_2017}. We model this process using an effective absorption cross-section of $\sigma_{\rm Lyman} = 5.23 \times 10^{-25} \: {\rm cm^{2}}$, based on \citet{wolcott-green_suppression_2011}.

\section{Initial conditions}
\label{sec:initial-conditions}

Following \citet{wollenberg_formation_2019}, we start our simulations from a simplified set of initial conditions that allow us to control the initial turbulent and rotational energy present in the gas. In a simulation box of size 13 pc we set up a Bonnor-Ebert sphere (BES) density profile. This has a density that is everywhere a factor of $f = 1.83$ larger than the density of a critical Bonnor-Ebert sphere, a central density $n_{\rm c} = 1.83 \times 10^{4} \: {\rm cm^{-3}}$, a radius $R = 1.87$~pc, a mass $M = 2671 \: {\rm M_{\odot}}$, and a uniform temperature $T = 200$~K. 

The gas is initially in solid body rotation and also has a random turbulent velocity component, generated as described in \citet{wollenberg_formation_2019}. The strength of the initial rotation and turbulence are parametrised by 
\begin{equation}
\beta_\mathrm{rot}=\frac{E_\mathrm{rot}}{|E_\mathrm{grav}|} \; \; \mathrm{and} \; \;  \alpha_\mathrm{turb}=\frac{E_\mathrm{turb}}{|E_\mathrm{grav}|}, 
\end{equation}
where $E_{\rm rot}$ is the rotational energy, $E_\mathrm{turb}$ is the turbulent kinetic energy and $E_\mathrm{grav}$ is the initial gravitational energy of the Bonnor-Ebert sphere. We carry out simulations using two different initial conditions, one turbulence-dominated ($\beta_{\rm rot} = 0.01, \alpha_{\rm turb} = 0.25$) and one rotation-dominated ($\beta_{\rm rot} = 0.1, \alpha_{\rm turb} = 0.001$). The value of $\beta_{\rm rot}$ adopted in our rotation-dominated setup is within the range of values found by \citet{hirano_one_2014} for Pop.~III star-forming clouds formed from cosmological initial conditions. Similarly, the value of $\alpha_{\rm turb}$ adopted in our turbulence-dominated setup -- which corresponds to a turbulent Mach number of order unity, given our choice of initial temperature -- is similar to the values encountered in cosmological simulations of Pop.~III star formation \citep[see e.g.][]{greif_simulations_2011}.

For each set of initial conditions, we run simulations using three different values of the sink accretion radius, $r_{\rm sink} = 2, 10,$ and 30~AU. Table~\ref{tab:init-settings} summarizes the different combinations of parameters used in the simulations.

\begin{table}
\begin{center}
\begin{tabular}{lcccc}
 Name & $r_\mathrm{sink}$ (AU) & $\beta_\mathrm{rot}$ & $\alpha_\mathrm{turb}$ & $t_\mathrm{coll}$ (kyr) \\ \hline
 T2   & 2  & 0.01 & 0.25  & 705.135 \\
 T10  & 10 & 0.01 & 0.25  & 716.665 \\  
 T30  & 30 & 0.01 & 0.25  & 716.665 \\  
 R2   & 2  & 0.1  & 0.001 & 743.686 \\
 R10  & 10 & 0.1  & 0.001 & 743.687 \\
 R30  & 30 & 0.1  & 0.001 & 743.687 \\
\end{tabular}
\caption{Summary of the initial settings of our simulations. $\alpha_\mathrm{turb}$ and  $\beta_\mathrm{rot}$ are ratios of the turbulent and the rotational kinetic energy to the gravitational energy, respectively. For each simulation we also list the sink accretion radius $r_\mathrm{sink}$ and the formation time of the first sink particle $t_\mathrm{coll}$.}
\label{tab:init-settings}
\end{center}
\end{table}

We follow the collapse of the whole BES until the first sink particle forms. This occurs around 700--750~kyr after the beginning of the simulation, depending on the adopted values of $\alpha_{\rm turb}$ and $\beta_{\rm rot}$ (see Table~\ref{tab:init-settings}). We temporarily halt the simulation and select a region of size 3.9~pc around the densest point. Subsequently, we continue to simulate this selected region using outflow boundary conditions. We do this for reasons related to how {\sc sprai} functions. As described in Section~\ref{sec:sprai} above, photon transport in {\sc sprai} occurs only on timesteps on which all the {\sc arepo} grid cells are synchronized. For reasons of accuracy, we do not want the synchronization timestep to be enormously larger than the natural hydrodynamical timestep of the densest cells. However, if we limit all the cells in our initial 13~pc box to a small timestep, this is computationally inefficient, as we then spend considerable time evolving cells far from the centre of the simulation on timesteps that are much shorter than their natural hydrodynamical timestep. Cutting out the central dense region and continuing the simulation with only this region mitigates this cost without significantly affecting the outcome of the simulation.  

Following the formation of the first sink particle, we run three variants of each simulation to help us better understand the effect of the radiative feedback on the surrounding gas. We denote them as follows:
\begin{description}
    \item[NF] -- no radiation feedback
    \item[RTP] -- with ionising radiation and radiation pressure
    \item[RTPr] -- as RTP, but with no absorption for $r < r_\mathrm{sink}$
\end{description}
The first two setups are self-explanatory, but the last deserves further comment. In setup RTPr, we modified {\sc sprai} to disable the absorption of photons within the accretion radius $r_\mathrm{sink}$ around all sink particles with masses higher than $M_\mathrm{sink}=10 \: {\rm M}_\odot$. Outside of $r_{\rm sink}$, the treatment of attenuation and photochemistry stay the same as in the usual version of the code. Radiation from sinks with masses $M < 10 \: {\rm M}_\odot$ is treated in the usual fashion. We discuss the motivation for this unusual setup in Section \ref{sec:rad-models-in-literature} below. 

\section{Results}
\label{sec:results}

All of our simulations show qualitatively similar behaviour prior to the onset of radiative feedback from massive ($M > 10 \: {\rm M_{\odot}}$) stars. As in many previous simulations of Pop.\ III star formation, H$_{2}$ cooling allows the gas to collapse quasi-isothermally over many orders of magnitude in density, resulting in the gas developing a power-law density profile. Eventually, a Pop.\ III protostar forms, surrounded by a dense accretion disk which soon thereafter fragments, yielding a compact cluster of interacting protostars. Some protostars are ejected by dynamical encounters, but those that remain close to the centre of the gas distribution
grow in mass over time through accretion from the surrounding gas reservoir. Once one or more stars reaches a mass of $M_{\rm star} \sim 10$--$20 \: {\rm M_{\odot}}$, they become hot enough to start to emit significant numbers of ionising and photodissociating photons (see Figure~\ref{fig:spectrum-temp-rad}; the precise mass at which this occurs depends on the accretion rate onto the star). From this point on, greater differences become apparent between the simulations, as we explore in more detail in the sections below.

\subsection{Collapse and fragmentation of the BES}

\begin{figure}
  \includegraphics[width=\linewidth]{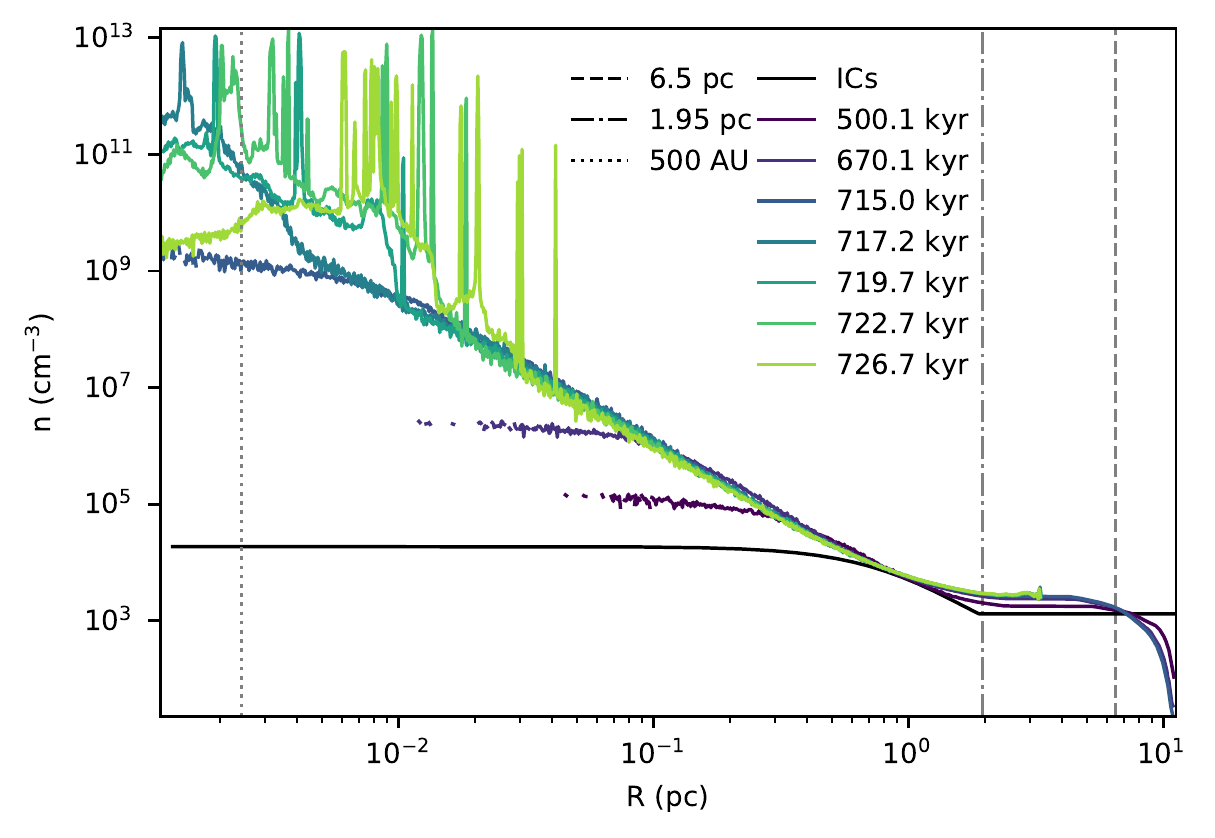}
  \caption{Spherically-averaged number density as a function of the distance from the centre of the collapse in simulation T10\_NF at various different output times. For reference, the first sink formed in this simulation at a time $t_{\rm coll} = 716.7$~kyr. Vertical lines denote the distance to the edge of the initial box, $R_\mathrm{box}=6.5$ pc (dashed), to the edge of the cutout region $R_\mathrm{cut}=1.95$ pc (dashed-dotted), and the inner 500~AU (dotted).}
  \label{fig:icsFigures-ndensProfile}
\end{figure}

\begin{figure}
  \includegraphics[width=\linewidth]{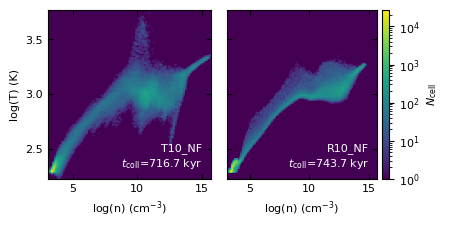}
  \caption{Density-temperature distribution at the onset of star formation in runs T10\_NF (left) and R10\_NF (right).}
  \label{fig:nT}
\end{figure}

As a representative example of the behaviour of the gas in the simulations during the initial collapse phase and subsequent onset of fragmentation, we show in Figure~\ref{fig:icsFigures-ndensProfile} the average gas number density as a function of the radial distance from the centre of mass in run T10\_NF at several output times. The gas first develops a power-law density profile with a slope of approximately -2.2, similar to the results found in previous simulations of Pop III star formation. After the formation of the first sink particle, which occurs at $t_\mathrm{coll}=716.7$~kyr, the whole system fragments into several high density regions that rotate around their common centre of mass and drift apart. Interactions between fragments are common, leading to preferential ejection of low mass sink particles, which often end up at very large distances from the centre of the simulation volume. A look at the density-temperature distribution of the gas (Figure~\ref{fig:nT}) prior to the onset of star formation shows several features that are familiar from previous studies of Pop III star formation: a steady temperature rise at $n > 10^{4} \: {\rm cm^{-3}}$, resulting from the inefficiency of H$_{2}$ cooling at densities greater than its critical density; a flattening in the $n$--$T$ relationship around $n \sim 10^{9} \: {\rm cm^{-3}}$ associated with the onset of three-body H$_{2}$ formation; and a subsequent temperature increase at high densities caused by the increasing optical depth of the H$_{2}$ ro-vibrational lines. The prominent spike in the temperature at $n \sim 10^{10} \: {\rm cm^{-3}}$ in run T10\_NF is the consequence of a shock occurring at this density, although it should be noted that this only involves a small fraction of the gas at these densities. Features such as this are more apparent in the turbulence-dominated runs and less apparent in the rotation-dominated ones, as demonstrated by the comparison between runs T10\_NF and R10\_NF in Figure~\ref{fig:nT}.

\begin{figure}
  \includegraphics[width=\linewidth]{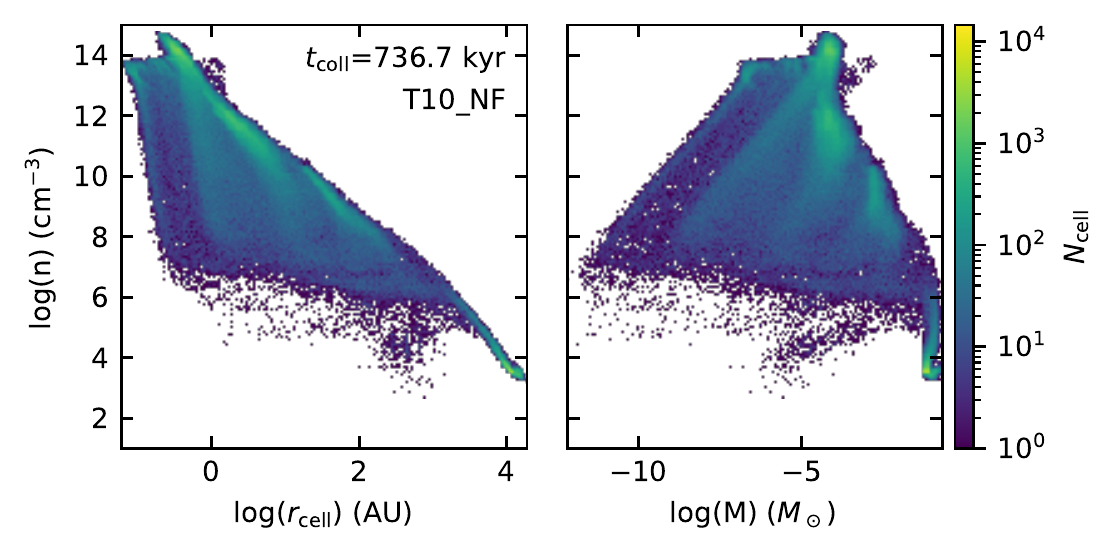}
  \caption{Distribution of cell radii (left) and masses (right) as a function of the density in the simulation T10\_NF at the final output time.}
  \label{fig:simulation-resolution}
\end{figure}

The top panels in Figure \ref{fig:simulation-resolution} illustrate the resolution achieved in the same simulation at the final output time, $t = 20$~kyr after the formation of the first sink particle. The left-hand panel in the figure shows the size distribution of the Voronoi mesh cells, quantified by their effective radii, $r_\mathrm{cell}= \left(3V/4\pi \right)^{1/3}$, where $V$ is the cell volume. The right-hand panel shows the distribution of the corresponding cell masses. Highly populated portions of the histogram correspond to different collapsing fragments. Similar resolutions are achieved in the other simulations. At densities characteristic of the central accretion disk ($n > 10^{11} \: {\rm cm^{-3}}$), all of the cells have sizes below 10~AU (with some being much smaller) and have masses $M_{\rm cell} < 10^{-3} \: {\rm M_{\odot}}$. At densities close to our sink formation threshold ($n \sim 10^{14} \: {\rm cm^{-3}}$),
all of the cells have radii $< 1$~AU and masses $M_{\rm cell} < 10^{-4} \: {\rm M_{\odot}}$. Note also that the absence of cells with $r_{\rm cell} \ll 1$~AU and densities $n > n_{\rm th}$ is a consequence of our sink particle algorithm: cells with radii this small lie inside the accretion radius of a sink and so accretion onto the sink therefore maintains their density at $n \leq n_{\rm th}$ by construction.

\begin{figure}
  \includegraphics[width=\linewidth]{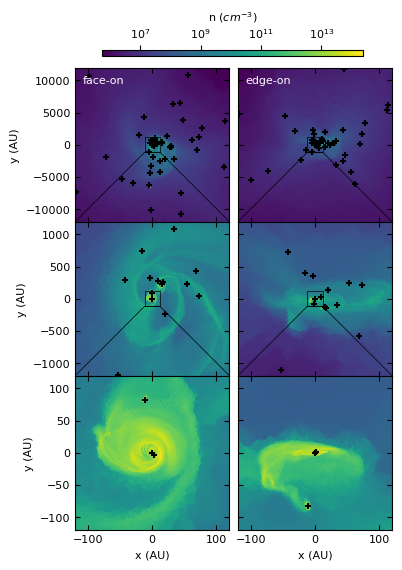}
  \caption{Face-on (left) and edge-on (right) projections of the column-averaged gas 
  number density in simulation T10\_NF centered on the most massive sink particle. Panels from top to bottom show box projections with sizes of 24000, 2400 and 240 AU, respectively. Crosses denote positions of the sink particles.}
  \label{fig:icsFigures-zoomProj}
\end{figure}

In Figure \ref{fig:icsFigures-zoomProj}, we show the positions of the sink particles in run T10\_NF at a time $t = 723.7$~kyr, roughly 7~kyr after the formation of the first sink. The rows from the top to bottom show face-on (left panels) and edge-on (right panels) column-averaged gas number density for boxes with sizes of 24000, 2400 and 240 AU, respectively. All of the plots are centered on the most massive sink particle present in the simulation, which remains close to the dynamical centre of the gas distribution. Black crosses denote the positions of the individual sink particles. The disk-like morphology of the gas distribution is plain, particularly in the lowest panel, as are the spiral arms created by the gravitational instability of the disk. The sinks all form within the disk, but interactions lead to many of the smaller sinks being ejected (see especially the upper panels).

\begin{figure}
  \includegraphics[width=\linewidth]{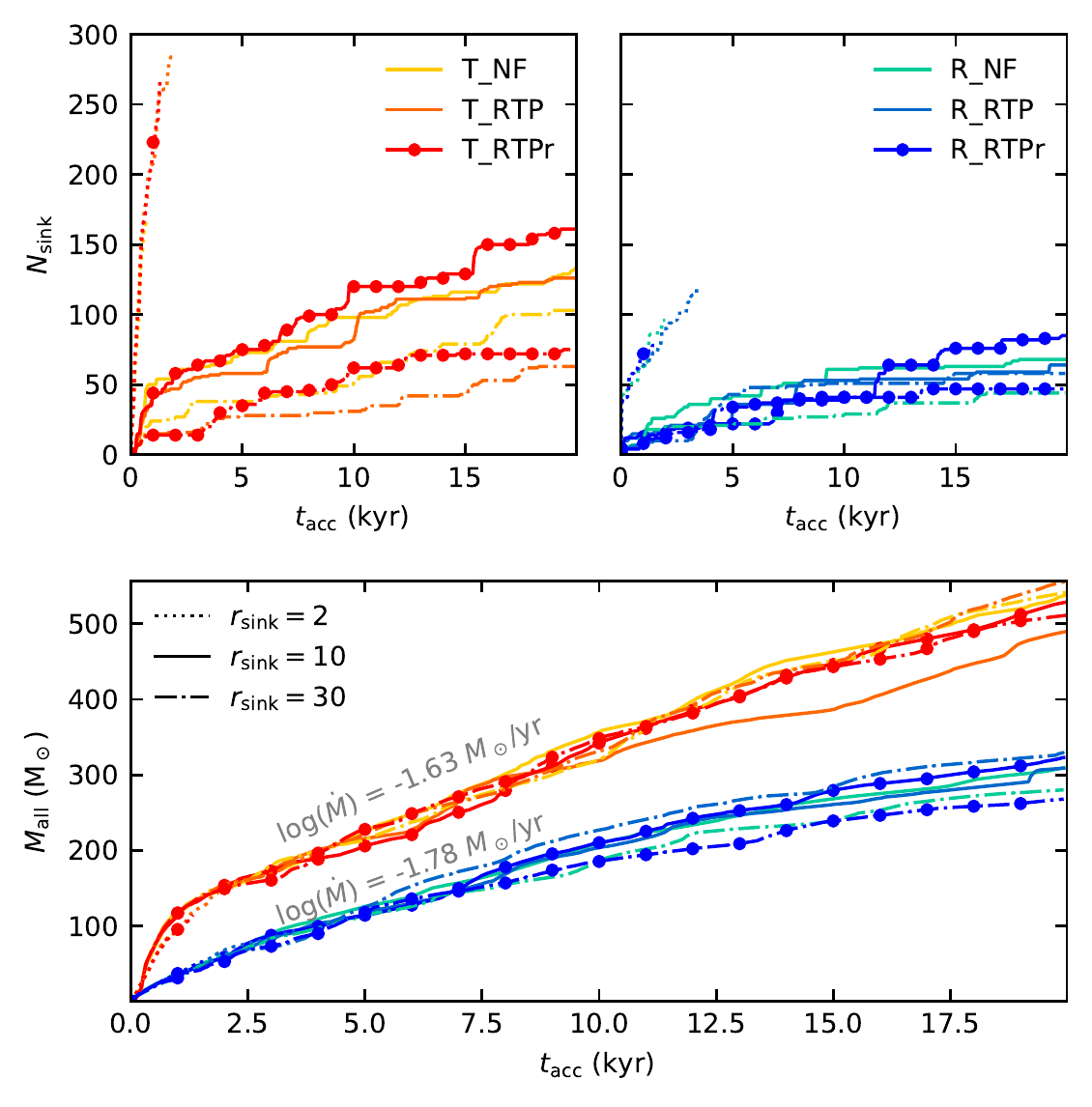}
  \caption{Time evolution of the total sink particle number (upper panels) and total mass accreted by sink particles (lower panel). In the lower panel we also indicate the average mass infall rates for the rotation-dominated and turbulence-dominated simulations.}
  \label{fig:sinks-numberOfSinks}
\end{figure}

If we look in more quantitative detail at the properties of the sink particles formed in the different simulations, differences become apparent. The panels in Figure~\ref{fig:sinks-numberOfSinks} show, from top to bottom, the time evolution of the number of sink particles and the total mass accreted by the sink particles, respectively. Note that the colour scheme and line styles used in Figure \ref{fig:sinks-numberOfSinks} are also used in later figures, to make it easy to distinguish the different simulations. Results from the simulations with high initial turbulence and low initial rotation are drawn in red tones, while those from the simulations with high initial rotation and low initial turbulence are shown in blue tones. Dots on the line plots indicate the runs in which absorption of radiation within $r_{\rm sink}$ is neglected.

As in \citet{wollenberg_formation_2019}, we find that the total mass accreted by the sink particles is sensitive to the amount of rotational energy in the initial conditions. Gas in the simulations with a higher initial $\beta_{\rm rot}$ collapses more slowly than in the runs with lower initial $\beta_{\rm rot}$, resulting in a systematically smaller total accretion rate and a difference in the total mass accreted of roughly a factor of two by $t = 20$~kyr. Amongst the runs with the same initial $\beta_{\rm rot}$, we find only minor variations in the total accreted gas mass, with no systematic trend with sink accretion radius or with the details of the radiative feedback. If we look at the number of sinks formed in each simulation, however, we find much greater variation, and in particular a strong systematic trend as a function of $r_{\rm sink}$. Runs with a smaller accretion radius form far more sink particles than those with a larger accretion radius. Indeed, so many sinks formed in the simulations with $r_{\rm sink} = 2$~AU that it became computationally prohibitive to keep the runs going for more than a few kyr. 
This high sensitivity to $r_{\rm sink}$ is a consequence of our sink particle algorithm, which does not allow new sinks to form within the accretion radius of existing sinks. This restriction is necessary to avoid artificial fragmentation of the gas, but means that we also miss some real small-scale fragmentation. That said, it is likely that many of the fragments forming within a few AU of each other would in reality merge \citep{wollenberg_formation_2019,susa_merge_2019} and so our runs with higher $r_{\rm sink}$ are arguably more representative of the real behaviour of Pop.~III star-forming systems.

\begin{figure}
  \includegraphics[width=\linewidth]{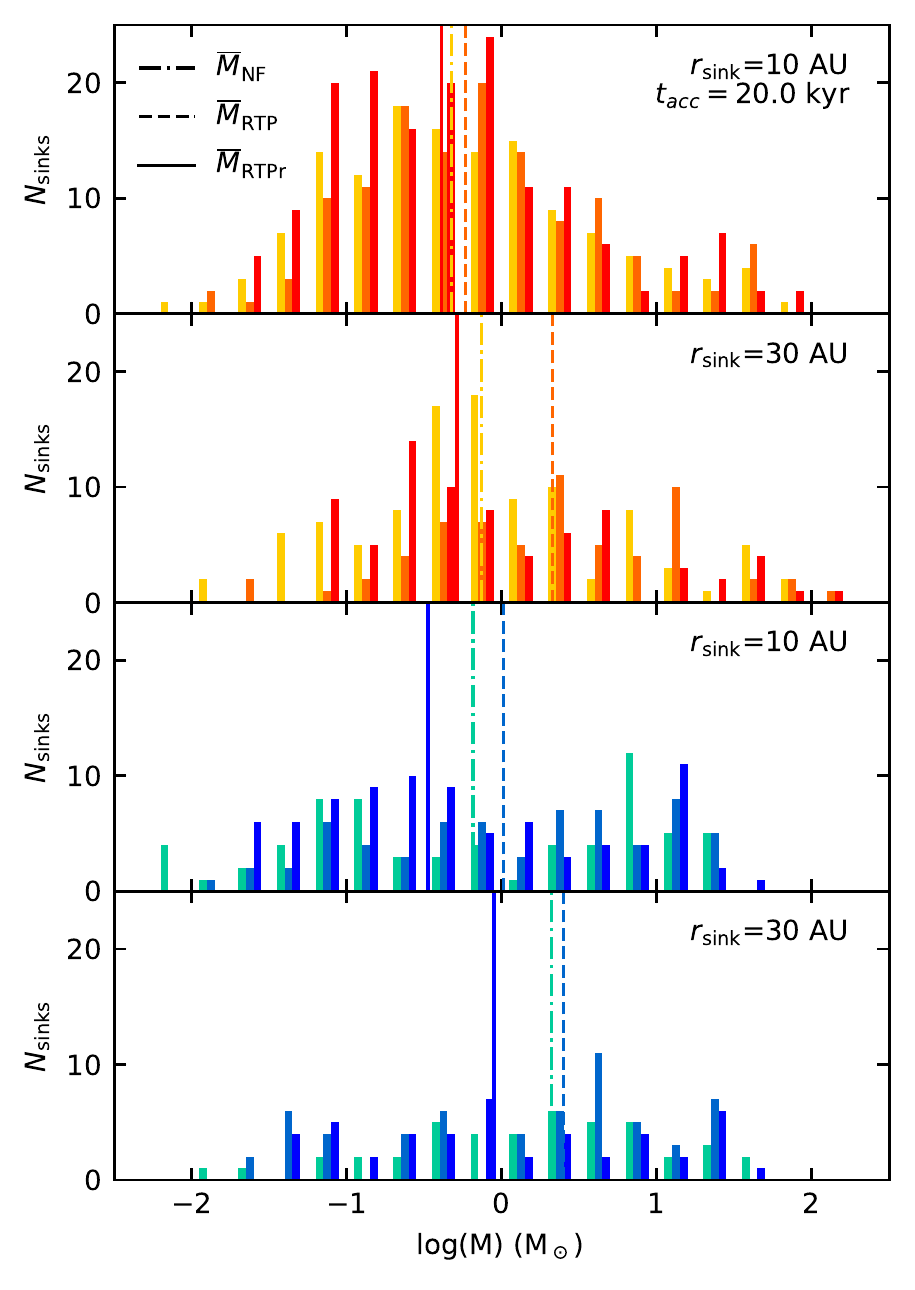}
  \caption{Mass distribution of sink particles in all our simulations that ran until time $t_\mathrm{acc}$=20 kyr. The colour scheme is the same as in Figure~\ref{fig:sinks-numberOfSinks}. Rows from top to bottom correspond to turbulent setups T10, T30 and rotational setups R10, R30, respectively. In each panel we show the results for our three different radiation setups: NF, RTP and RTPr. The vertical dotted lines denote the median values for each distribution.}
  \label{fig:sinks-massFunction}
\end{figure}

We continued all of the simulations with sink accretion radii of 10 or 30 AU to a final time $t_\mathrm{acc} =20$~kyr after the formation of the first sink. During this period, each simulation produced between 40 and 150 sink particles. In a recent study, \citet{susa_merge_2019} argue that previous simulations of Pop.\ III star formation find similar amounts of fragmentation if compared using a suitable scale-free time $\tau =  \sqrt{4 \pi G \rho_{\rm th}}$, where $\rho_{\rm th}$ is the density at which sink particles are created. In our case, his model predicts that the simulations should form around 10 sinks by $t = 20$~kyr, around an order of magnitude smaller than the number we actually find. Our simulations therefore do not agree with the \citet{susa_merge_2019} prediction. On the other hand, we do find that at times $> 1$~kyr, the number of sinks grows roughly as $N_{\rm sink} \propto t^{0.3}$, in good agreement with the scaling predicted by \citet{susa_merge_2019}.

We show the final mass distributions of the sinks in the different simulations in Figure~\ref{fig:sinks-massFunction}. In this figure, the rows from top to bottom correspond to setups T10, T30, R10 and R30, respectively. Three different bar plots in each row show our three radiation setups NF, RTP and RTPr, respectively. The median values (dotted lines) for each distribution are also indicated in each plot. 

Looking at the mass distributions, we see that the runs with high initial turbulence produce a clear peak between 0.1 and 1~M$_\odot$, although overall the distribution remains much flatter than we find for the present-day stellar initial mass function \citep[see e.g.][]{kroupa_initial_2002}. On the other hand, the runs with high initial rotation do not display any clear peak. Minor differences are apparent in the mass distributions recovered from simulations with different radiation setups, but the significance of these variations is unclear as they are comparable to the run-to-run variations found by \citet{wollenberg_formation_2019} for simulations differing only in the random realization of the initial turbulent velocity field. 

To summarize: the main quantities determining how much fragmentation occurs in the simulations and how much mass is accreted by sinks are the initial values of $\alpha_{\rm turb}$ and $\beta_{\rm rot}$ and the sink accretion radius $r_{\rm sink}$, with the latter affecting only the amount of fragmentation and not the total accretion rate. On the other hand, the details of the radiative feedback have very little impact on either the number of sinks formed or the total mass accreted, at least on a 20~kyr timescale. In the next section, we explore why this is the case.

\subsection{Radiation feedback around sinks}
\label{sec:rad-feedbac-around-sinks}

In this section, we analyze in more detail the environment surrounding selected sink particles. As Figure~\ref{fig:sinks-massFunction} demonstrates, most of the sinks that form in the simulations have masses $M < 10 \: {\rm M_{\odot}}$ and hence do not produce a significant number of ionising photons. We therefore focus here on the most massive sinks, which we might expect to have the greatest impact on their surroundings. Specifically, we show results for the 
most massive sinks in each simulation.
Moreover, we show results only for the runs with $r_{\rm sink} = 10$ or 30~AU, as the runs with $r_{\rm sink} = 2$~AU did not form any high mass sink particles by the time we ended the simulations. 

\begin{figure*}
  \includegraphics[width=\linewidth]{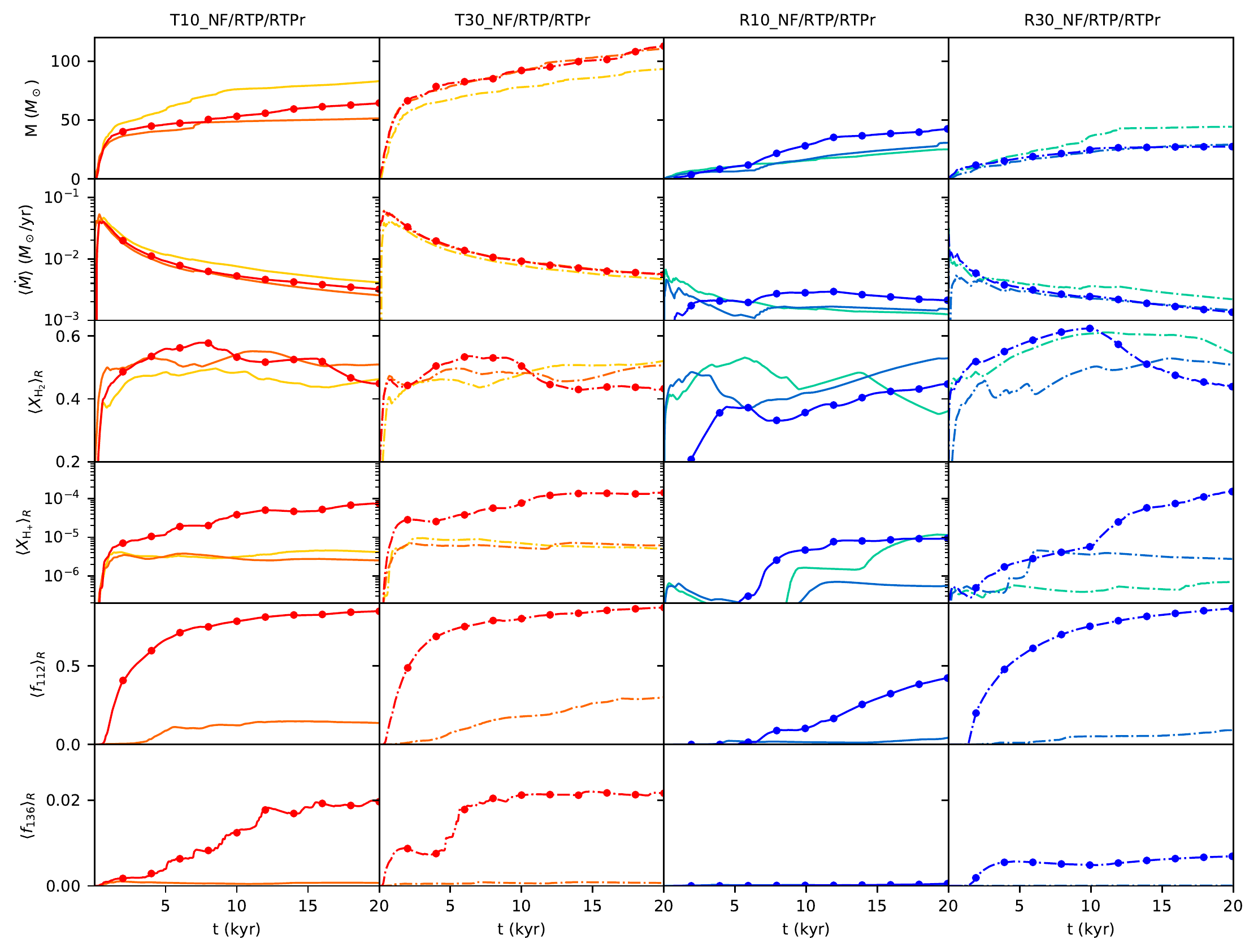}
  \caption{Evolution of gas properties close to the most massive protostars. In each simulation, we select the 
  most massive sink particle and compute various properties within a sphere of 200~AU centered on it. Columns from left to right correspond to simulations T10, T30, R10 and R30. In each column we compare three different radiation setups: NF, RTP and RTPr. 
  Rows from the top to the bottom correspond to masses $M$, accretion rates $\dot{M}$, mass fractions $X$ of H$_2$ and H$^+$ and volume filling fractions $f_{112}$ and $f_{136}$ of radiation in the 11.2+ and 13.6+ energy bins, respectively. Lines in this figure represent the time evolution of the properties above. The meaning of the different line styles and colours is the same as in Figure~\ref{fig:sinks-numberOfSinks}.}
  \label{fig:sinks-evolution}
\end{figure*}

Figure \ref{fig:sinks-evolution} summarizes the time evolution of several selected properties of the sink particles and the region surrounding them. Each column corresponds to a different simulation setup: T10, T30, R10 or R30, indicated in the label at the top. In every column we plot the results for our three different radiation setups: NF, RTP and RTPr.

The gas around the sink particles is often dynamically unstable and changes on short timescales. When sink particles pass through different environments, orbit around their binary companions or fly close to other sink particles, values such as their accretion rate or the chemical composition of the gas surrounding them can vary rapidly, making it difficult to directly compare results from different simulations in a meaningful fashion. In practice, we have found that it becomes much easier to compare the different simulations if we compare the cumulative time-weighted average values of various properties rather than their instantaneous values. These average values are computed as
\begin{equation}
    \langle \xi \rangle=\Bigg\{\sum_0^{j=i} \xi_j \frac{(t_j-t_{j-1})}{t_i} \Bigg\}_{i=0}^{N},
    \label{eq:ct-average}
\end{equation}
where $N$ is the total number of snapshots output by the code after the formation of the first sink particle\footnote{Even though the radiation field is updated on intervals smaller than 1 year, snapshots are produced only every 10 years.}, $i$ and $j$ are the indices of individual snapshots, $t_{i}$ and $t_{j}$ are the times of snapshots $i$ and $j$ (measured from the formation of the first sink particle, with $t_{-1}\equiv0$), and $\mathcal{\xi}_j$ is the value of the property of interest $\mathcal{\xi}$ in snapshot $j$. The properties averaged in this way can be properties of the sink (e.g.\ its mass or accretion rate) or of the gas surrounding the sink. In the latter case, we compute spatially averaged values within a sphere of radius $R = 200$~AU surrounding the sink particle and denote them as $\langle \xi \rangle_R$.

The first and second rows in Figure \ref{fig:sinks-evolution} show masses $M$ and corresponding mean accretion rates $\langle \dot{M} \rangle$ of the selected sink particles. We see that accretion onto the sinks is more effective in the runs with high initial turbulence and low initial rotation (high $\alpha_{\rm turb}$, low $\beta_{\rm rot}$) than in those with low 
initial turbulence and high initial rotation (low $\alpha_{\rm turb}$, high $\beta_{\rm rot}$), in agreement with the behaviour of the total accretion rate onto all sinks. There is also a tendency for the sinks with larger accretion radii to also have larger accretion rates. Furthermore, we see that in most cases, the sink accretion rate declines over time. Most of the sinks we examine here gain the majority of their mass within the first 1-2~kyr of the simulation and thereafter increase their mass only slowly with time. This behaviour is very similar regardless of whether or not we include the effects of radiative feedback, implying that it is the turbulent dynamics of the gas surrounding the sinks that largely determines how much gas they can accrete and not the radiative feedback from the sinks themselves. 

The third and fourth rows show the average mass fractions of molecular hydrogen $\langle X_\mathrm{H_2} \rangle_R$ and ionised hydrogen $\langle X_\mathrm{H^+} \rangle_R$ in the 200~AU region surrounding each sink. The mass fraction of atomic hydrogen is not shown, but is trivial to calculate since $X_{\rm H} = 1 - X_{\rm H_{2}} - X_{\rm H^{+}}$. We see that in every case, the gas in the region surrounding each massive sink is dominated by H and H$_{2}$, with only a small amount of H$^{+}$ present. This is to be expected in the runs without radiative feedback, but we find essentially the same behaviour in the runs with radiative feedback. In other words, the ionising radiation produced by the massive stars is unable to ionise a significant fraction of the surrounding gas. This is a key result of our study, but appears to be at odds with the results of many previous studies of ionising feedback from Pop.\ III stars. 

To help us understand this behaviour, we have examined the average volume filling fraction of the photons in the 11.2+ and 13.6+ bins, $\langle f_{112} \rangle_R$ and $\langle f_{136} \rangle_R$, defined as the fraction of the volume of the 200~AU sphere that contains any photons in these energy bins. These filling fractions are illustrated in the fourth and fifth rows of the Figure \ref{fig:sinks-evolution}.\footnote{Although not shown here, the ionising radiation in the higher energy bins behaves in a very similar fashion to that in the 13.6+ bin.} We see straight away that in the RTP runs, almost no ionising radiation escapes from the immediate vicinity of the stars and only a small fraction of the volume is exposed to Lyman-Werner photons. On the other hand, in the RTPr runs, where absorption of radiation close to the sink is disabled, Lyman-Werner photons fill a large fraction of the volume of the 200~AU sphere, although ionising photons are still confined to a relatively small volume. 
Taken together, these results suggest that in our fiducial RTP runs, the HII regions and PDRs surrounding the massive stars are trapped in the dense accretion disk and never penetrate into the lower density gas above and below the disk. On the other hand, in the RTPr runs, the lack of absorption very close to the stars allows radiation to escape from the disk, although most of the dense gas in the disk -- representing the majority of the mass in the 200~AU sphere -- remains unaffected by this radiation. We will examine this interpretation of our results in more detail in the next section.

\subsection{HII region trapping}
\label{sec:comparison-radiation}
\subsubsection{Spatial distribution of the radiation and ionised gas}
\label{sec:spatial-dist}

\begin{table}
\begin{center}
\begin{tabular}{cclc}
 Mass ($M_\odot$) & $r_\mathrm{centre}$ (AU) & Simulation & Age (kyr) \\ \hline
 67.78 &   0    & T30\_RTP  & 19.78 \\
 46.17 &   2.68 & T30\_RTP  & 19.86\\
 13.81 &  74.54 & T30\_RTP  & 5.25\\
  5.38 &  13.9  & T30\_RTP  & 4.86\\
\hline
 68.66 &   0    & T30\_RTPr & 19.86\\
 49.33 &   2.25 & T30\_RTPr & 19.78\\
 36.83 &  41.2  & T30\_RTPr & 16.93\\
  1.03 & 187.72 & T30\_RTPr & 0.62\\
 \end{tabular}
\caption{Summary of sink particle properties in the 400 by 400 AU region analyzed in Section \ref{sec:comparison-radiation}. The radius $r_\mathrm{centre}$ is measured from the location of the most massive sink particle in the region. Values are shown at a time $t_{\rm acc} = 20 \: {\rm kyr}$ after the formation of the first sink particle.}
\label{tab:sinks-summary}
\end{center}
\end{table}

To better understand why UV photons in the RTP simulations are prevented from escaping the accretion disk, we analyze the gas properties in the immediate vicinity of some of the massive stars formed in the simulations. Specifically, we focus on the second and third most massive sink particles that form in simulation T30\_RTP. As we discuss later, the behaviour of this system is representative of the behaviour we see for all of the massive stars in the RTP simulations. By the end of the simulation, these two sinks are located in a tight binary with a separation of $\sim 2$~AU. A similar pair of sinks with similar masses and separation is found in our T30\_RTPr simulation, allowing us to directly compare the behaviour of the gas and the radiation field in both cases.

In our simulations, both massive stars in the binary emit large numbers of ionising and photodissociating photons into their common environment. In principle, it would be possible to examine the effects of the radiation from each star individually. However, given the small separation between the two stars, it is easier to consider their combined radiation field.

\begin{figure}
  \includegraphics[width=\linewidth]{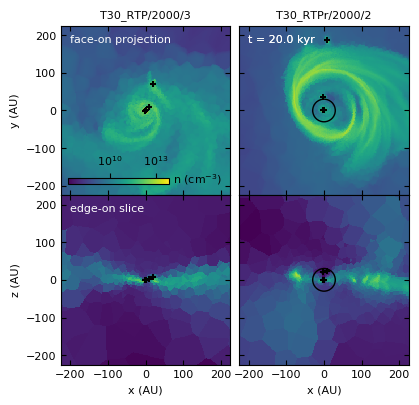}
  \caption{Face-on projection (top) and edge-on slices (bottom) of the gas number density around two selected sink particles (columns) at time $t_\mathrm{acc}$=20 kyr. In case of the projections we plot the column-averaged number density. Small black crosses mark the positions of the main and companion sink particles within the selected volume.}
  \label{fig:mainSink-detailDensity}
\end{figure}

\begin{figure}
  \includegraphics[width=\linewidth]{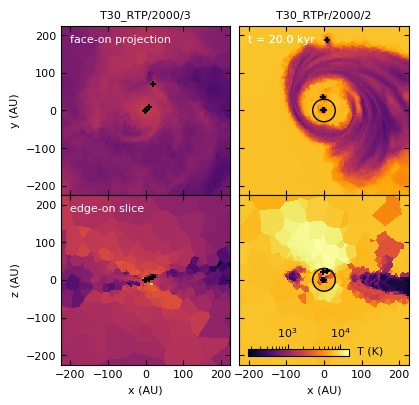}
  \caption{As Figure~\ref{fig:mainSink-detailDensity}, but showing the temperature of the gas. In the case of the projections, this is the column-averaged value.}
  \label{fig:mainSink-detailTemperature}
\end{figure}

In Figure~\ref{fig:mainSink-detailDensity}, we show plots of gas number density as a face-on projection (top row) and edge-on slice (bottom row) of a 400 by 400~AU region centred on the most massive sink in the binary system in simulations T30\_RTP and T30\_RTPr at a time $t_{\rm acc} = 20 \: {\rm kyr}$. Figure~\ref{fig:mainSink-detailTemperature} shows similar plots of the temperature structure of the gas. In both Figures, we indicate the location of the binary sink particles using black crosses. For completeness, we also indicate the locations of the other sink particles present in this region. The masses and ages of these sinks are summarized in Table~\ref{tab:sinks-summary}, along with their distance from the most massive sink in the region, $r_{\rm centre}$. 

From the figures, we see immediately that the massive stars in the T30\_RTP simulation are embedded in the centre of an extended accretion disk with a temperature of $\sim 1000$~K and a density which peaks in the centre. A disk is also present in the T30\_RTPr simulation, but with a clear difference in morphology: there is a lower density cavity close to the location of the binary which is associated with a region of hot ($T \sim 10^{4}$~K) gas that is particularly apparent above the disk midplane.

\begin{figure*}
  \includegraphics[width=0.8\linewidth]{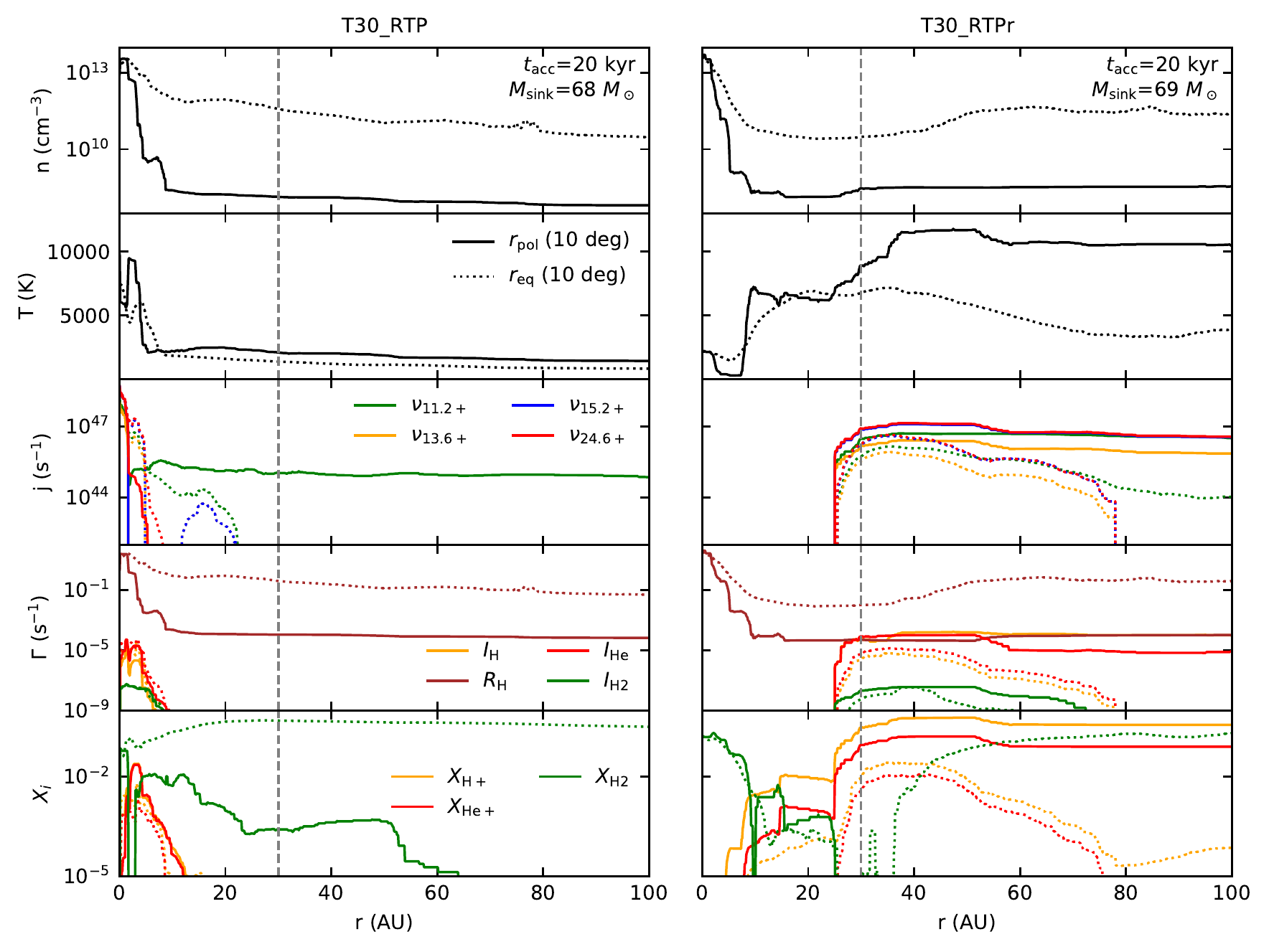}
  \caption{Mean polar $r_\mathrm{pol}$ and equatorial $r_\mathrm{eq}$ radial profiles around the most massive sink particle in the region studied in Section~\ref{sec:comparison-radiation} in simulations T30\_RTP (left column) and T30\_RTPr (right column), calculated as described in the text. The results are shown for a time $t_\mathrm{acc} = 20$~kyr. We indicate the corresponding masses $M_\mathrm{sink}$ of the sink particles in the top panels. Profiles in the panels from the top to bottom show gas density, temperature, photon fluxes in all four radiation bins ($\nu_\mathrm{11.2+}$, $\nu_\mathrm{13.6+}$, $\nu_\mathrm{15.2+}$ and $\nu_\mathrm{24.6+}$), recombination $R_\mathrm{H}$ and ionisation rates ($I_\mathrm{H}$, $I_\mathrm{He}$ and $I_\mathrm{H_2}$) and mass fractions of different gas species ($X_\mathrm{H^+}$, $X_\mathrm{He^+}$ and $X_\mathrm{H_2}$), respectively. The dashed grey lines indicate the accretion radius of the sink particle.}
  \label{fig:mainSink-diskRays}
\end{figure*}

In Figure \ref{fig:mainSink-diskRays} we plot mean radial profiles of various properties around the most massive member of the binary in the simulations T30\_RTP (left column) and T30\_RTPr (right column) at time $t_\mathrm{acc}=20$ kyr. To measure these profiles, we chose a set of 300 rays and distributed their orientations equally on a $r=100$ AU sphere around the sink particle using the {\sc healpix} algorithm \citep{gorski_healpix_2005,zonca_healpy_2019}. We then calculated the local properties of the gas along each ray. We next divided the whole solid angle into two regions according to the angle $\theta\in(0, 180^\circ)$ measured with respect to the angular momentum vector of the accretion disk. We averaged values along rays with $80^{\circ}<\theta<100^{\circ}$ to obtain mean values as a function of equatorial distance~$r_\mathrm{eq}$, and did the same for rays with $\theta<10^{\circ}$ and $\theta > 170^{\circ}$ to obtain mean values as a function of polar distance $r_\mathrm{pol}$. Although it is a fairly crude representation of the real 3D complexity of the region, this procedure nevertheless allows us to distinguish between the dense gas in the accretion disk (which extends approximately $10^{\circ}$ from the equatorial plane) and the much lower density gas in the polar regions. We also note that although we only show the results for the most massive member of the binary, the results for profiles centred on the other member will be very similar at radii much greater than  the binary separation of 2--3~AU.

\begin{figure}
  \includegraphics[width=\linewidth]{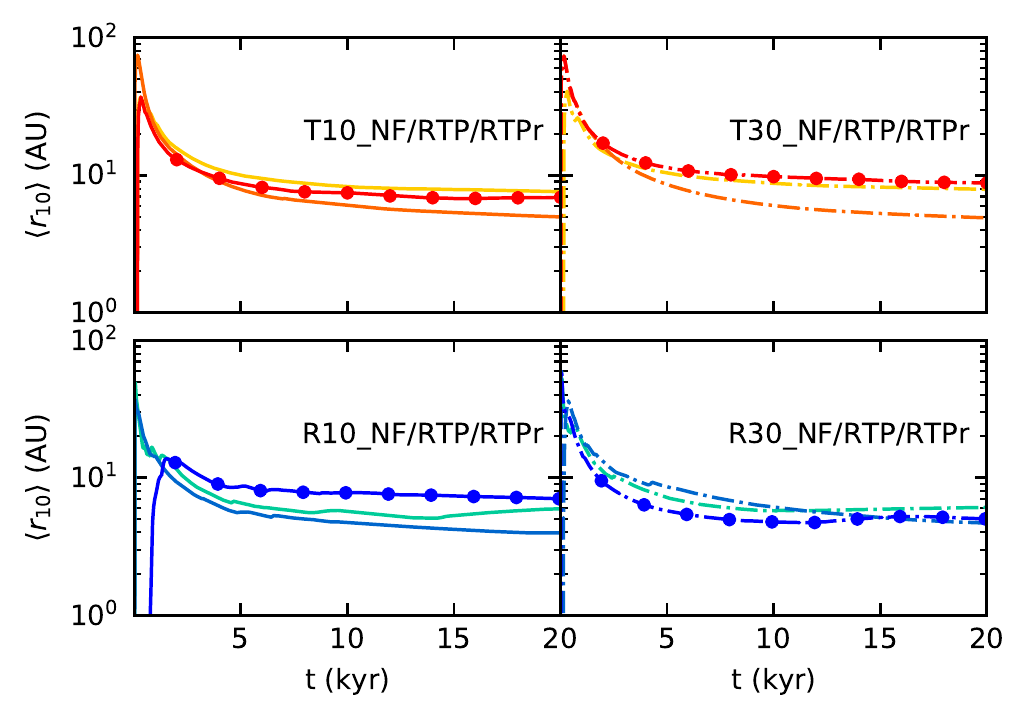}
  \caption{Distance of the sink particles from Figure \ref{fig:sinks-evolution} to the nearest cell with number density below $n=10^{10} \: {\rm cm^{-3}}$. The distances are averaged in time according to Equation \ref{eq:ct-average}.}
  \label{fig:sinks-evolDisk}
\end{figure}

The top panels in Figure~\ref{fig:mainSink-diskRays} show the mean density profile in the equatorial and polar directions in runs T30\_RTP and T30\_RTPr. Since the sink particles are located in a flattened disk, it is unsurprising that we find an anisotropic density distribution surrounding them. In the equatorial direction, the density first falls off rapidly from a few times $10^{13} \: {\rm cm^{-3}}$ close to the sink to around $10^{12} \: {\rm cm^{-3}}$ at $r_{\rm eq} \sim 10$~AU, but thereafter decreases only slowly with increasing equatorial distance. On the other hand, in the polar direction the density decrease close to the sink is far more pronounced, with $n$ dropping to $\sim 10^{9} \: {\rm cm^{-3}}$ by the time that $r_{\rm pol} = 20$~AU.  

To verify that this behaviour is not simply due to the fact that we have selected some special time in the simulation, we have calculated the distance from each of our considered sinks to the nearest gas cell with density $n < 10^{10}$ cm$^{-3}$ at various output times. (We discuss this choice of density threshold in Section \ref{sec:radiation-vs-density} below). The results are shown in Figure~\ref{fig:sinks-evolDisk}. Initially, this distance -- which we can take to be a reasonable proxy for the disk thickness -- is large, but after only a few kyr, it settles down to a value of $\sim 10$~AU or smaller for most of the sinks, consistent with the behaviour we have already seen in Figure~\ref{fig:mainSink-diskRays} for the most massive sinks. Importantly, this means that in our T10, R10, T30 and R30 runs, the sink accretion radius is comparable to or larger than the thickness of the accretion disk near the sinks. 

\begin{figure}
  \includegraphics[width=\linewidth]{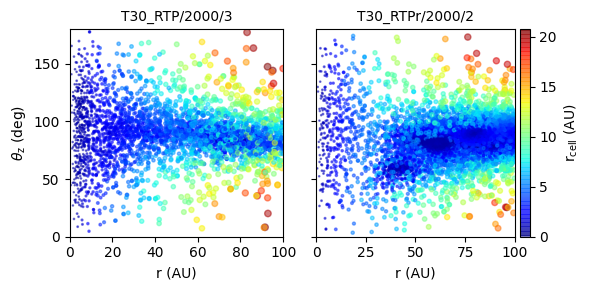}
  \caption{Radii of the grid cells close to the most massive sink at $t_{\rm acc} = 20$~kyr in simulations T30\_RTP (left) and T30\_RTPr (right). The locations of the cells are shown using a polar coordinate system centred on the sink and the colour-coding indicates the cell radius $r_{\rm cell}$.}
  \label{fig:cellsizes}
\end{figure}

We have also checked that the disk thickness that we recover is not simply due to our numerical resolution, i.e.\ that we are not simply finding an unresolved disk with a thickness of one or two Voronoi cells. In Figure~\ref{fig:cellsizes}, we show the radii of the cells close to the most massive sink in runs T30\_RTP and T30\_RTPr, plotted in polar coordinates. In the disk and close to the sink, the cells have radii of at most 1--2~AU, consistent with the values that we would expect given their densities (see Figure~\ref{fig:simulation-resolution}). Close to the sink, the disk thickness is around 20~AU and so the disk is resolved in the vertical direction with around 5-10 cells. Although not large, this number of cells should be sufficient to resolve the pressure gradient in the vertical direction, and hence we are confident that the value we recover for the disk thickness is physically meaningful and not simply a numerical artifact.

Turning to the temperature structure of the gas surrounding the massive stars, we see that in run T30\_RTP, there is clear evidence of radiative heating close to the star, but that once we move more than a few AU away, this vanishes and the temperature drops to the value of a few 1000~K that is characteristic of the disk even in the absence of radiative heating. In run T30\_RTPr, on the other hand, the temperature rises as we move away from the sink, reaching a value of 6000--7000~K in the equatorial plane and $\sim 10^{4}$~K in the polar regions. 

\begin{figure}
  \includegraphics[width=\linewidth]{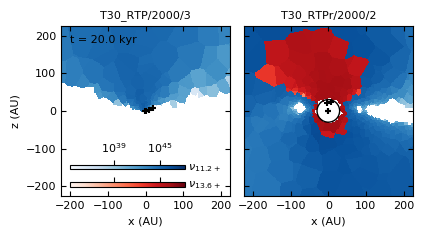}
  \caption{Edge-on slices of the radiation flux in the 11.2+ eV (blue) and 13.6+ eV (red) bins around the same two sinks as Figure \ref{fig:sinks-evolution}.}
  \label{fig:mainSink-detailFlux}
\end{figure}

The panels in the third row in Figure~\ref{fig:mainSink-diskRays} show that this difference in temperature structure is the result of a clear difference in the spatial distribution of the photon fluxes. In run T30\_RTP, radiation is emitted directly from the sink particle location and the ionising photon flux is completely attenuated within a distance of 20~AU from the sink, with most of the attenuation occurring within\footnote{Note that the secondary flux peak around 15~AU from the centre is caused by the other member of the binary.} the central 10~AU. Therefore, gas close to the star is heated by the radiation but the majority of the gas remains unaffected. Only the photons in the 11.2+ bin are able to escape from the immediate surroundings of the massive star, and even in that case, escape is only possible in the polar direction. In run T30\_RTPr, the behaviour of the radiation is very different. In this run, we effectively inject the photons at the accretion radius of the sink, and from there they can efficiently propagate through the surrounding gas, producing substantial radiative heating. To help us better visualize the behaviour of the radiation field in the two runs, we show in Figure \ref{fig:mainSink-detailFlux} an edge-on slice of the photon fluxes in bins 11.2+ eV (blue) and 13.6+ eV (red) in each run. We overlay the two flux maps on top of each other. In both cases there is a preferential radiation outburst direction towards positive z values. However, in run RTP, only the photons in the 11.2+ eV bin escape, while in run RTPr we see significant escape of ionising radiation, indicative of the formation of an HII region with a size of $\sim 200$~AU.

In the fourth row in Figure~\ref{fig:mainSink-diskRays}, we plot the ionisation rates of H, H$_{2}$ and He, along with an estimate of the maximum recombination rate per hydrogen nucleus, 
\begin{equation}
    R_\mathrm{H} = \alpha_\mathrm{B}(T) n_\mathrm{H},  \label{eq:rec}
\end{equation}
i.e.\ the rate per H nucleus at which the gas would recombine were it to be fully ionised. Here $\alpha_\mathrm{B}(T)$ is the case B recombination coefficient \citep{1938ApJ....88...52B}. The true recombination rate is of course a factor of $\sim X_{\rm H^{+}}^{2}$ smaller than this, where $X_{\rm H^{+}}$ is the mass fraction of ionised hydrogen. We see that in the RTP simulation, the ionisation rates of the three species close to the source are several orders of magnitude smaller than the maximum recombination rate, implying that the gas in this region should have a low fractional ionisation. Far from the source, the discrepancy is even more pronounced. In the RTPr simulation, however, the ionisation rate in the polar regions matches the maximum recombination rate for $r_{\rm pol} > 30$~AU, i.e.\ as soon as we inject the ionising photons. We would therefore expect the gas in the polar direction to be highly ionised.

\begin{figure}
  \includegraphics[width=\linewidth]{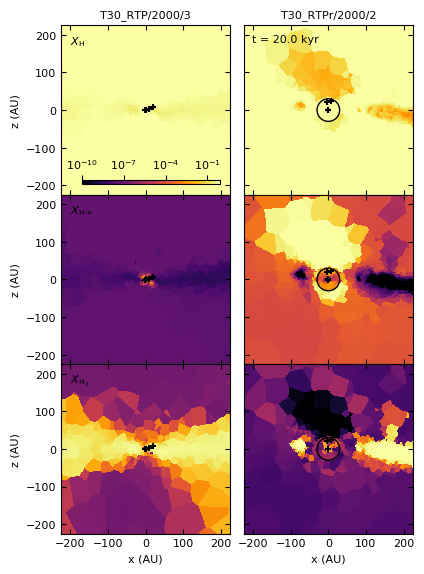}
  \caption{Edge-on slices of the neutral (top), ionised (middle) and molecular (bottom) hydrogen mass fractions around the same two sinks as Figure \ref{fig:sinks-evolution}. We clipped mass fraction values in this figure into the interval (10$^{-10}$, 1) to gain a better contrast.}
  \label{fig:mainSink-detailAbund}
\end{figure}

Our expectations based on the ionisation and recombination rates are borne out when we look at the actual spatial distributions of the ionised ($X_\mathrm{H^+}$ and $X_\mathrm{He^+}$) and molecular ($X_\mathrm{H_2}$) gas (bottom row in Figure~\ref{fig:mainSink-diskRays}). In simulation T30\_RTP, the fractional ionisation outside of the sink accretion radius is very small and the molecular gas fraction remains comparatively high. In simulation T30\_RTPr, however, dissociation of H$_{2}$ and photoionisation are far more effective, leading to low H$_{2}$ mass fractions in the polar direction, coupled with a high ionised gas fraction.

In Figure~\ref{fig:mainSink-detailAbund}, we show image slices of different hydrogen species around our chosen sink particles. A significant difference between the RTP and RTPr setups is the amount of molecular hydrogen in the accretion disk. The disk of the former is symmetric and almost entirely composed of H$_2$. The latter disk is disrupted, especially at the sink radius of the binary and high H$_2$ fractions are located only in the densest parts of the disk, typically the spiral arms. There is also a clear difference in the ionisation state of the gas in the two runs, with run T30\_RTPr showing evidence for a distinct HII region above the disk. Comparing this Figure with the temperature slice shown in Figure~\ref{fig:mainSink-detailTemperature} and the radiation flux slice shown in Figure~\ref{fig:mainSink-detailFlux}, we see that there is, as we would expect, a good correlation between the region with high ionising photon flux and the hot, highly ionised gas.

Finally, while the trapping of the HII region provides a simple explanation for why radiative feedback has no appreciable impact on the accretion of mass by the sink particles in simulation T30\_RTP, the reader might reasonably ask why we also see little impact of feedback on the accretion rate in simulation T30\_RTPr (see Figure~\ref{fig:sinks-evolution}). The reason for this is that accretion onto the sinks takes place primarily through the accretion disk, and the dense gas in the disk is not strongly affected by the growth of the HII region and PDR even in the simulation where they are not trapped. A similar result has been found in previous studies of radiative feedback from Pop.\ III stars \citep[see e.g.][]{hosokawa_formation_2016}, which show that the way in which feedback reduces and eventually terminates accretion is by disrupting the inflow of lower density gas onto the accretion disk. This occurs once the HII region has reached a size of $\sim 10000$~AU, and hence corresponds to a later evolutionary stage than simulation T30\_RTPr has reached by the time at which we halt it. It is likely that if we were to continue this simulation for considerably longer, we would also start to see the influence of the feedback on the accretion rate.

\subsubsection{Why is the HII region trapped near the stars?}
\label{sec:radiation-vs-density}
The analysis presented in Section~\ref{sec:spatial-dist} above shows that in run T30\_RTP, the ionising photons produced by the massive binary that we examined are unable to escape from the dense accretion disk surrounding the binary. Here, we examine with the aid of some simple quantitative models why this is the case. 

\begin{figure}
  \includegraphics[width=\linewidth]{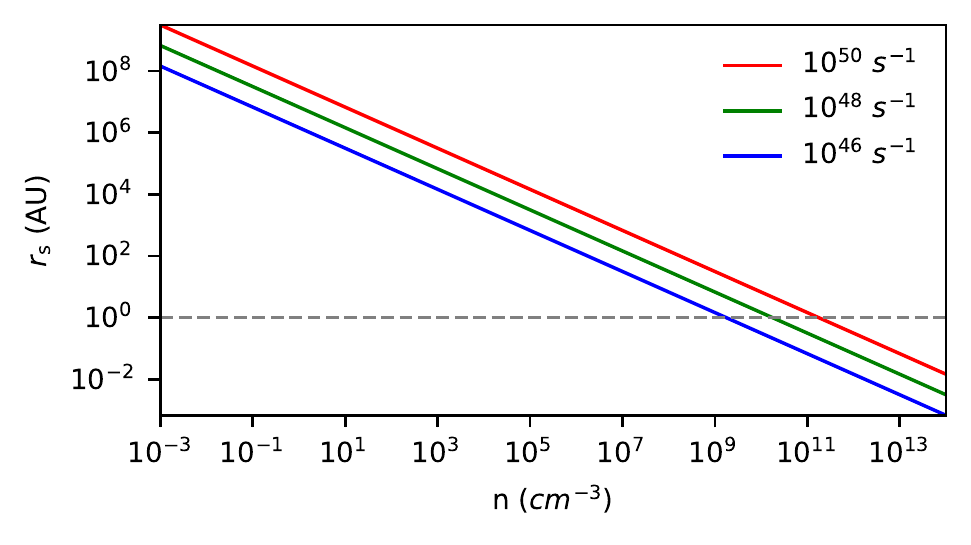}
  \caption{Str\"{o}mgren radius $r_\mathrm{s}$ as a function of the number density $n$ for three different ionisation sources $\dot{Q}$ (coloured lines). The grey dashed line indicates a radius of 1~AU, comparable to the size of the majority of the cells in the vicinity of the sink particle (see Figure~\ref{fig:simulation-resolution}). Str\"omgren spheres with radii much smaller than this value are not resolved in our simulations.}
  \label{fig:icsFigures-radThermPress}
\end{figure}

A simple starting point is the Str\"omgren radius. A point source of radiation emitting $\dot{Q}$ ionising photons per second embedded in uniform density hydrogen\footnote{The inclusion of helium changes $r_{\rm s}$ by only a small amount.} with number density $n_{\rm H, 0}$ will rapidly produce an HII region with radius
\begin{equation}
    r_{\rm s} = \left( \frac{3 \dot{Q} }{4\pi \alpha_\mathrm{B} n_\mathrm{H,0}^2} \right)^{1/3}, \label{eq:const_rho}
\end{equation}
where $\alpha_{\rm B}$ is the case B recombination coefficient. The time taken for an HII region to reach this radius is approximately given by the recombination time $t_{\rm rec} = (n_{\rm H, 0} \alpha_{\rm B})^{-1}$, which is of the order of a few years or less in the dense gas close to the massive stars formed in our simulation. Figure \ref{fig:icsFigures-radThermPress} shows how $r_{\rm s}$ evolves with density for $\dot{Q} = 10^{46}$, $10^{48}$ and $10^{50} \: {\rm photons \, s^{-1}}$. We see that even for $\dot{Q} = 10^{50} \: {\rm s^{-1}}$ -- corresponding to the number of ionising photons emitted by a $120 \: {\rm M_{\odot}}$ star \citep{schaerer_properties_2002} -- the Str\"omgren radius becomes smaller than an AU for number densities $n > 10^{11} \: {\rm cm^{-3}}$, characteristic of the gas in the accretion disk. Therefore, in the dense environment of the disk, we expect the initial size of the HII region to be smaller than the disk scale height.\footnote{Note that although Equation~\ref{eq:const_rho} is formally valid only for uniform density gas, the fact that the value it yields for $r_{\rm s}$ is much smaller than the disk scale height retrospectively justifies the use of a constant density approximation.}

The gas in the HII region has an equilibrium temperature of $\sim 10^{4} \: {\rm K}$ and hence is over-pressured compared to the surrounding gas. If the gravitational attraction of the central ionising source can be neglected, it is easy to show that this elevated pressure will drive the hydrodynamical expansion of the HII region, leading to a power-law dependence of the HII region size on time in the case of approximately uniform gas \citep{spitzer_lyman_physical_1978,hosokawa_dynamical_2006-1} or a so-called champagne flow in the case of a steeply stratified density distribution \citep[see e.g.][]{franco_formation_1990,shu_self-similar_2002}. Close to the star, however, its gravitational attraction cannot be neglected and the HII region can expand hydrodynamically only if the sound speed of the ionised gas exceeds the escape velocity associated with the gravitational field of the star \citep{keto2002}. In our example case of a massive binary, the escape velocity of the larger of the two stars is given by
\begin{equation}
v_{\rm esc} = \left(\frac{2GM_{*}}{r}\right)^{1/2} \simeq 350 \, {\rm km \, s^{-1}} \left(\frac{M_{*}}{68 \: {\rm M_{\odot}}}\right)^{1/2} \left(\frac{1 \: {\rm AU}}{r}\right)^{1/2},
\end{equation}
where $r$ is the distance from the star and where the first term in brackets is of order unity. This becomes equal to the ionised gas sound speed at a radius known as the Bondi-Parker radius, given by \citep{keto_formation_2007}
\begin{equation}
r_{\rm bp} = \frac{2 G M_{*}}{c_{\rm s, i}^{2}},
\label{eq:bp}
\end{equation}
where $c_{\rm s, i}$ is the sound speed of the ionised gas. If we take the temperature of the ionised gas to be approximately 15000~K (see Figure~\ref{fig:mainSink-detailTemperature}), then for our example case of a $68 \: {\rm M_{\odot}}$ star, this yields we have $r_{\rm bp} \simeq 350$~AU, a distance considerably larger than the initial Str\"omgren radius. Therefore, thermal pressure is unable to expand the HII region significantly beyond its initial size.

\begin{figure}
  \includegraphics[width=\linewidth]{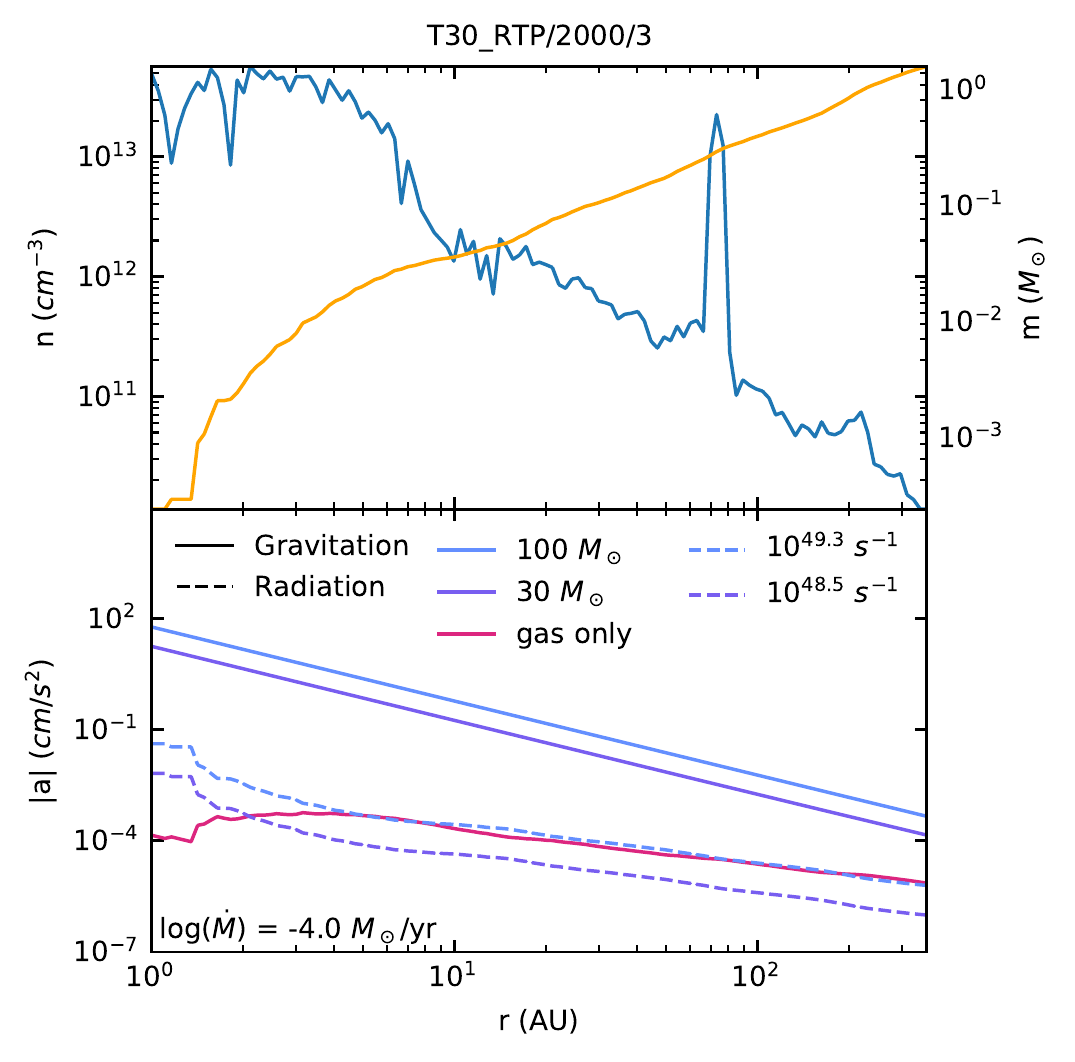}
  \caption{The top panel shows spherically averaged number density (blue) and enclosed mass (orange) as a function of the distance from the most massive sink particle in the simulation  T30\_RTP at time 20 kyr (see Table \ref{tab:sinks-summary}). The pronounced peak in the density at $\sim 80$~AU is due to a neighbouring sink particle. For simplicity, we exclude the mass of this object in our calculation of $a_{\rm G}$. The bottom panel shows absolute values of the gravitational acceleration $a_{\rm G}$ (solid lines), and the acceleration caused by photon ionisation per hydrogen atom $a_{\gamma}$ (dashed line). The magenta solid line indicates the gravitational acceleration due to the gas alone, while the other solid lines denote the combined effect of the gas and a central star of the specified mass as illustrative examples. The two dashed lines show the corresponding radiative acceleration. Note that it is always significantly smaller. Note also that the mass of the central star in our fiducial model is $\sim 68 \: {\rm M_{\odot}}$, and so the corresponding values fall between the blue and purple curves.}
  \label{fig:icsFigures-gravPotential}
\end{figure}

It is also straightforward to show that the inclusion of the effects of radiation pressure does not change this basic conclusion. In the top panel of Figure \ref{fig:icsFigures-gravPotential}, we plot spherically-averaged radial profiles of gas density (blue) and enclosed gas mass (orange) around sink particle 3 in the simulation T30\_RTP at a time of 20 kyr. In the bottom panel of the Figure, we compare the gravitational acceleration at various distances from the star with the acceleration resulting from the absorption of ionising photons. We compute the gravitational acceleration following
\begin{equation}
a_\mathrm{G} = \frac{G}{r^2}\int^r_0  4 \pi r'^2 \rho (r') dr' + \frac{G M_*}{r^2},
\end{equation}
where $\rho(r)$ is the spherically-averaged radial density function of the gas, $G$ is the gravitational constant and $M_{*}$ is the mass of the star. To highlight the influence of the stellar gravity, we show profiles for two different values of the central stellar mass ($M_*={30, 100}$ M$_\odot$), plus one case where the star is absent. We see that on scales smaller than a few hundred AU, the gravitational acceleration is dominated by the contribution from the star.

The dashed lines in Figure \ref{fig:icsFigures-gravPotential} show the mean radiative acceleration within a sphere of radius $r$ around the star, computed assuming that all of the ionising photons are absorbed within this radius, and plotted as a function of $r$. This is computed following
\begin{equation}
a_\gamma = \dot{Q} \frac{E_\gamma}{c} \frac{1}{M(r)},
\end{equation}
where $E_{\gamma}$ is the mean energy of the ionising photons, and $M(r)$ is the mass enclosed within $r$,
\begin{equation}
M(r) = \int_{0}^{r} 4\pi r'^2 \rho(r') dr'.
\end{equation}
$a_\gamma$ is the radiative acceleration that the gas would feel if the radiation pressure force was distributed uniformly within the HII region. Strictly speaking, this is not the case: the force acting on a given parcel of gas within the HII region depends on its recombination rate and hence will be higher in denser gas. However, as we expect that any ionised gas that does feel a stronger force will interact with and transfer momentum to the surrounding ionised gas, the mean radiative acceleration is the best way to quantify the response of the HII region as a whole to the radiation pressure.

Figure \ref{fig:icsFigures-gravPotential} shows results for two different ionising photon production rates, $\dot{Q}={10^{48.5},10^{49.3}}$ photons per second and a mean photon energy $E_\gamma=15$~eV. These ionising photon production rates correspond to the values produced by stars of 30 and 100 M$_\odot$, respectively, assuming a fixed accretion rate of $\dot{M} = 10^{-4} \, \mathrm{M_\odot \, yr^{-1}}$. As seen from the plot, the radiative acceleration is highest when the HII region is very small and drops as the HII region increases in size. However, we also see that, regardless of our choice of HII region size, the radiative acceleration exceeds the gravitational acceleration only in the case where $M_{*} = 0$, i.e.\ when we do not account for the gravity of the star producing the ionising photons. In the more realistic case where $M_{*} > 10 \: {\rm M_{\odot}}$, we see that even at 1~AU, the gravitational acceleration exceeds the radiative acceleration by more than three orders of magnitude. We can therefore safely conclude that radiation pressure will also be unable to drive the hydrodynamical expansion of the HII region, which therefore remains trapped in the dense accretion disk.

To summarize: the high gas density in the accretion disk surrounding the massive stars limits the initial size of the Str\"omgren radius to $< 1$~AU. This close to the star, its gravitational attraction is stronger than the forces acting on the gas due to thermal pressure or radiation pressure, and so the HII region is unable to expand hydrodynamically and instead remains trapped in the disk.

\subsection{Comparison of all sink particles}
The number of sink particles formed in our simulations is relatively large and so it is impractical to examine all of them in as much detail as we have done for the example in the previous section. Instead, we have attempted to analyze in a more automated fashion the properties of the gas surrounding each of the massive sink particles formed in our simulations.

\begin{figure*}  \includegraphics[width=\linewidth]{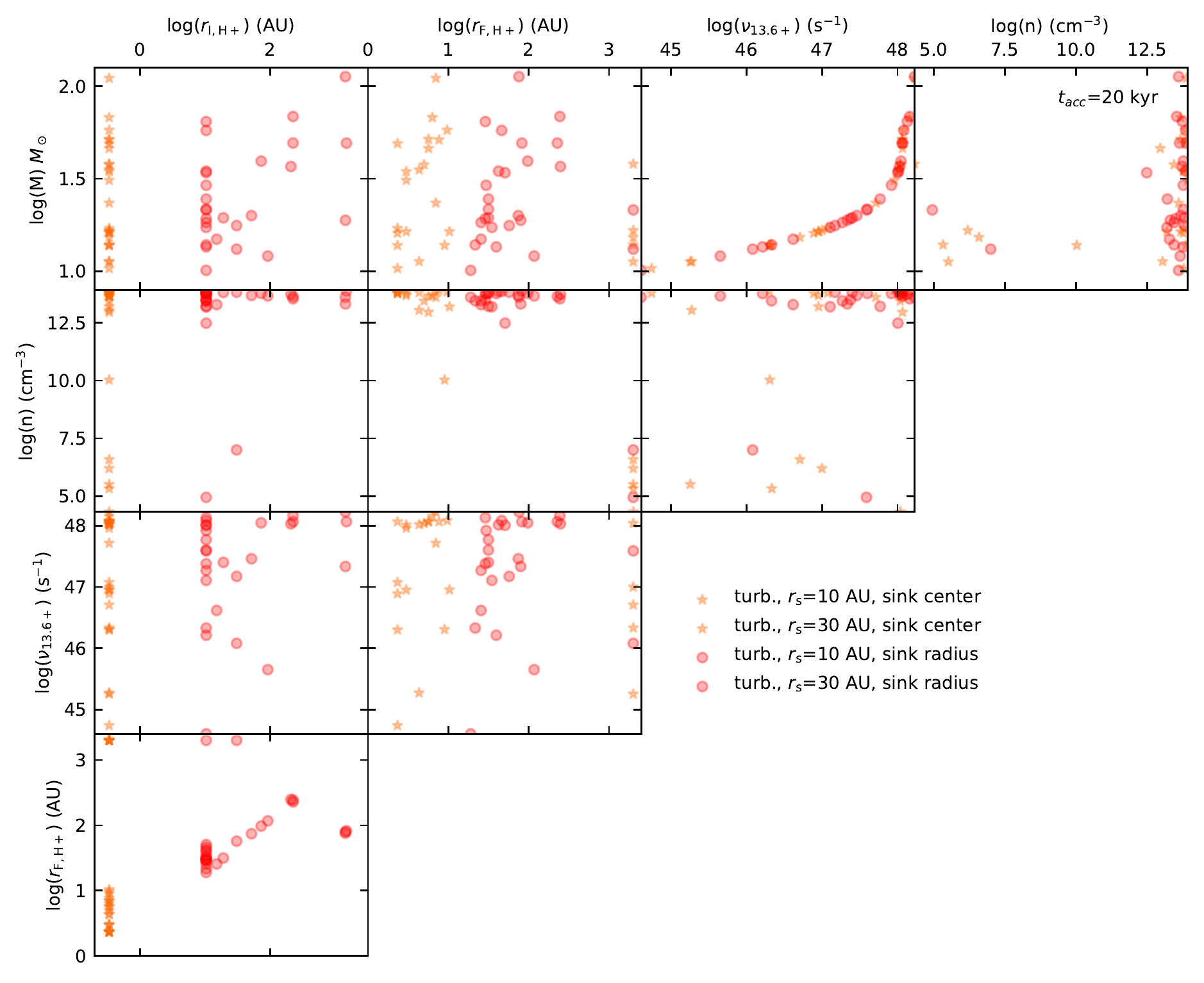}
  \caption{Correlation matrix showing the relationship between the sink particle mass $M$, emission rate of ionising photons in the 13.6+ bin $\nu_{13.6+}$, 
  gas density at the sink location $n$ and extent of the region affected by ionising radiation. The latter is calculated in two ways: $r_\mathrm{I,H+}$ is the size of the region where the fractional ionisation exceeds 80\% and $r_\mathrm{F,H+}$ is the size of the region where the flux of ionising photons is greater than zero. The colour scheme adopted here is the same as in Figure \ref{fig:sinks-numberOfSinks}. Stars denote results from our RTP runs, which account for attenuation within the sink accretion radius. Circles denote results from the RTPr runs, where attenuation within the sink accretion radius is artificially suppressed.}
  \label{fig:sinks-correlations}
\end{figure*}

Figure \ref{fig:sinks-correlations} is a correlation matrix showing the correlations between five
important quantities that we have measured for each sink particle: the sink mass $M$, the local number density $n$, the emission rate of ionising photons in the 13.6+ bin ($\nu_{13.6+}$)
and two different measures of the size of the region affected by ionising radiation from the sink, $r_\mathrm{F,H+}$ and $r_\mathrm{I,H+}$. The first of these measures, $r_\mathrm{F,H+}$, denotes the largest distance from the sink at which the flux of ionising photons is non-zero, while the other, $r_\mathrm{I,H+}$, denotes the length of the longest continuous radial ray that one can draw from the source within gas which has an ionisation fraction $X_{\rm H^{+}} > 0.8$. (Note that these definitions imply that the smallest possible value for $r_\mathrm{F,H+}$ is the size of the grid cell containing the sink particle, since that cell will always contain some ionising photons). All six quantities shown here were computed for the final output time of the simulation, $t_\mathrm{acc}=20$~kyr. For clarity, only the values for stars with masses $> 10 \: {\rm M_{\odot}}$ are shown, since lower mass stars produce an insignificant amount of ionising radiation. Due to the gravitation interactions with their more massive companions these stars are ejected far (>$10^4$ AU) from the accretion regions. In general, runaway stars exist in our simulations, but the masses are not enough to cause significant photoionisation.

We see immediately from this correlation plot that the behaviour we analyzed in detail in the previous section is common to most of the massive stars. In our RTP runs (denoted by orange 
stars in the Figure), none of the stars are able to produce an HII region large enough to escape from the immediate surroundings of the star. This is also true in many of the RTPr runs, although in this case the minimum size of the HII region is set by the sink accretion radius, explaining the vertical lines of points visible at 10 and 30~AU in the panels on the left of the plot. The only stars that do succeed in producing HII regions with sizes $> 100$~AU are a small number of stars with masses $M > 30 \: $M$_{\odot}$ in our RTPr runs (including the massive binary system analyzed previously). Here, the high ionising fluxes of the stars plus the suppression of attenuation within the sink accretion radius combine to allow the stars to create bipolar HII regions that can expand to large distances. 

The reason that most of the stars do not create sizeable HII regions is easy to understand when we look at the density of the gas in which they are situated. In most cases, this is extremely high, and so the stars can only produce HII regions with sizes that are small compared to the local scale-length of the dense gas. The only exceptions are a small number of stars with masses 10--$20 \: {\rm M_{\odot}}$ that have been ejected from the central star-forming region and that lie at a distance of $\sim 10^{4} \: {\rm AU}$ from any other sink. In this case the problem is the low ionising fluxes produced by these stars, which do not allow them to produce large HII regions despite the low ambient densities. 

Finally, we can also see from the plot that for most stars, the size of the region around the star that contains a non-zero ionising photon flux is generally larger than the size of the highly ionised HII region. This is expected behaviour and is a consequence of our finite grid resolution: as previously noted, the minimum size of the region with non-zero flux is one grid cell, and 
grid cells located just outside of the ionisation front will typically also have very small but formally non-zero fluxes. Nevertheless, it is clear that even if we use this more conservative definition of the region affected by each massive star, the size of these regions is still very small, with the largest barely exceeding 100~AU.

In conclusion, our analysis shows that the massive stars formed in our RTP simulations cannot produce and maintain HII regions with sizes larger than a grid cell, and that this is a general feature of the simulation and not a specific peculiarity of the binary system examined in detail in Section~\ref{sec:comparison-radiation}. Sizeable HII regions are only produced if we artificially suppress the attenuation of ionising radiation close to the stars, as in our RTPr simulations.

\section{Discussion}
\label{sec:discussion}

\subsection{Comparison with previous work}
\label{sec:rad-models-in-literature}
\subsubsection{Simulations of Pop.\ III star formation}

Simulations of Pop.\ III star formation that have sufficient resolution to follow the build up of the protostellar accretion disk and that also account for the impact of the ionising photons produced by the stars that form are highly computationally expensive. For this reason, only a handful of such studies have been carried out. In Table~\ref{tab:popiii-radiation-attempts}, we summarise the key details from a representative sample of these studies. 

\begin{table*}
\begin{tabular}{lccccc}
 Study & $r_\mathrm{sink}$ (AU) & $n_\mathrm{max}$ (cm$^{-3}$) & Photons
 inserted & Method \\ 
 & & & at sink boundary & & \\ \hline
 \citet{hosokawa_protostellar_2011} &  12 & 10$^{12}$ & Yes & 2D nested grid \\
 \citet{stacy_first_2012} &  50 & 10$^{12}$ & Yes & SPH  \\
 \citet{susa_mass_2013} &  30 & $3\times10^{13}$ & Yes & SPH \\
 \citet{susa_mass_2014} &  30 & $3\times10^{13}$ & Yes & SPH \\
 \citet{stacy_building_2016} & 1 & $10^{16}$ & No, but see text & SPH \\
 \citet{hosokawa_formation_2016} & 30 & $10^{13}-10^{14}$ & Yes & 3D spherical grid \\  
 \citet{sugimura_birth_2020} & 64 & $2\times10^{11}$ & No & AMR \\
 This paper & 10, 30 & $7.2\times10^{13}$ & No & Moving mesh \\
 \hline
\end{tabular}
\\
\caption{List of previous high-resolution simulations of radiative feedback from Pop.\ III stars. The parameter $r_{\rm sink}$ is the size of the sink cell or the accretion radius of the sink particle used to represent each star in each simulation and $n_{\rm max}$ is the density at which these are created. Unless otherwise noted, all studies were carried out in 3D and tracked both ionising and LW photons.}
\label{tab:popiii-radiation-attempts}
\end{table*}

These previous studies have generally found that far ultraviolet radiation readily escapes from the vicinity of the disk, creating a large photodissociation region (PDR) in the polar directions. In most cases, ionising radiation also easily escapes and drives the expansion of a bipolar HII region above and below the disk.\footnote{The simulations presented by \citet{susa_mass_2013} and \citet{susa_mass_2014} are an exception here: in these simulations the HII region does not manage to break out of the disk during the period studied.} The effect of the PDR (and the HII region, if present) is to slow down the infall of fresh gas onto the protostellar accretion disk. At a somewhat later time, the disk itself is photoevaporated. This behavior stands in sharp contrast to our finding that the ionising radiation and most of the Lyman-Werner band radiation remains trapped in the dense accretion  disk close to the star, never reaching the lower density surrounding gas. 

Why is the result we obtain so different from those obtained by these previous studies? As in our own study, these simulations made use of sink cells or sink particles to represent massive Pop.\ III stars. However, a common feature of the majority of these simulations is that they injected ionising radiation into the simulation at the edge of the region represented by the sink: at the sink cell boundary in the case of the grid codes, or at the accretion radius in the case of most of the smoothed particle hydrodynamics (SPH) simulations. In other words, their treatment of UV radiation was very similar to that in our RTPr models. However, as we have already seen, simulations in which the photons are injected at the edge of the sink accretion radius produce qualitatively different results from those in which the photons are injected at the actual location of the massive star (our RTP models). In most of these studies and in our own RTPr models, the size of the sink is larger than the scale height of the protostellar disk in the vicinity of the star, and so by injecting radiation at the edge of the sink, one artificially bypasses the dense gas in the disk, allowing the HII region to expand into the lower density gas above and below the disk. Importantly, this can allow the ionising radiation to reach gas which has a density low enough that the Str\"omgren radius exceeds the Bondi-Parker radius, with the result that the ionised gas can easily escape from the gravitational influence of the central star, allowing the HII region to expand to large distances.
On the other hand, if the radiation is injected into the simulation at the actual location it is produced, the HII region it creates is trapped and never escapes into this lower density gas.

As far as we are aware, only two previous 3D studies of ionising feedback from Pop.\ III stars have considered the case where ionising radiation is injected into the simulation at or very close to the massive star: \citet{stacy_building_2016} and \citet{sugimura_birth_2020}. In \citet{stacy_building_2016}, the propagation of ionising radiation from the most massive Pop.\ III star formed in their simulation is modelled by interpolating the density, temperature and ionisation fraction from their SPH particle distribution to a logarithmically-spaced spherical grid and then tracing rays on this grid. The size of the central grid cell was set to 1~AU, meaning that in principle this approach should be able to correctly capture the initial stages of the interaction between the HII region and the disk. However, \citet{stacy_building_2016} note that the resolution of their SPH simulation is too low to allow this. They therefore assume that the earliest stages of the expansion of the HII region are well described by the self-similar champagne flow solution of \citet{shu_self-similar_2002}, switching to a self-consistent calculation of the HII region growth only once the radius of the HII region exceeds 200~AU. The problem with this approach is that the \citet{shu_self-similar_2002} solution only accounts for the self-gravity of the ionised gas and not for the gravitational attraction of the massive star. As we have already seen, this is not a good approximation on scales smaller than $\sim 100$~AU and hence the \citet{shu_self-similar_2002} solution does not correctly describe the dynamics of the ionised gas on these scales.

In \citet{sugimura_birth_2020}, ionising radiation is traced along rays starting at $r_{\rm sink}$, but an effort is made to account for the effects of absorption within $r_{\rm sink}$. For each direction that is traced with a ray, \citet{sugimura_birth_2020} calculate the size of the initial Str\"{o}mgren radius, $r_{\rm s, init}$ using the density just outside of $r_{\rm sink}$ in that direction as a proxy for the density inside of $r_{\rm sink}$. In directions where $r_{\rm s, init} < r_{\rm sink}$, no ionising photons are injected, while in directions with $r_{\rm s, init} > r_{\rm sink}$, photons are injected at the sink radius with no attenuation (K.~Sugimura, private communication). This scheme does a better job of capturing the interaction between the ionising radiation and the dense accretion disk than in previous studies. However, the large value used for the sink accretion radius ($r_{\rm sink} = 64$~AU) means that in the polar directions, the density used to compute $r_{\rm s, init}$ is the density of the gas some considerable distance above the accretion disk. As this is much lower than the density of the gas in the disk, \citet{sugimura_birth_2020} find that  $r_{\rm s, init} > r_{\rm sink}$ in those directions, which would not be the case if the density used to calculate $r_{\rm s, init}$ was instead the characteristic density close to the centre of the accretion disk, as in our RTP runs. We therefore expect that if \citeauthor{sugimura_birth_2020} were to repeat their calculation using a much smaller value for $r_{\rm sink}$, they would recover results very similar to those in our simulations. 

\subsubsection{Simulations of present-day star formation}
Although the possibility that HII regions will be trapped around massive Pop.\ III stars has attracted little previous attention (with the exception of the early 1D study by \citealt{omukai2002} and a brief discussion in \citealt{susa_mass_2014}), it has been understood for a considerable time that the same phenomenon may occur during the formation of present-day massive stars. \citet{mestel1954} was the first to point out that if accretion onto a massive star were sufficiently rapid, the HII region produced by the star would be unable to escape from its immediate vicinity, despite the elevated pressure of the ionised gas compared to the surrounding neutral gas, and a number of more recent analytical studies have come to very similar conclusions \citep{walmsley1995,keto2002,keto2003,Krumholz2018}. This topic has also been studied numerically by \citet{sartorio2019}, who confirm that in some circumstances the HII region can be completely trapped by the surrounding accretion disk. On the other hand, HII region trapping is not observed in the recent high resolution simulations of \citet{Kuiper2018} because the bipolar outflow driven by the central source clears away enough gas in the polar directions to allow the HII region to escape. 

\subsection{The importance of high resolution}
Our analysis of HII region trapping in Section~\ref{sec:comparison-radiation} allows us to address an important technical question: what numerical resolution is required in simulations of Pop.\ III star formation in order to have a chance of observing this phenomenon? An obvious starting point is the Bondi-Parker radius (Equation~\ref{eq:bp}). If this is not resolved, then trapping of the HII region becomes impossible, since the thermal pressure of the gas within it will always exceed the gravitational attraction of the ionising source, meaning that the HII region will always expand. 

An additional condition comes from the requirement that we resolve the structure of the gas up to the density at which the local value of the Str\"omgren radius becomes equal to the Bondi-Parker radius, i.e.\ $r_{\rm s} = r_{\rm bp}$. The relevant density depends on the stellar mass, the ionising photon flux and the temperature of the ionised gas, but for reasonable choices for these parameters is typically around $n \sim 10^{8} \: {\rm cm^{-3}}$. In practice, this will often be the more restrictive requirement, since, as we have already seen, the gas density drops off rapidly above and below the protostellar accretion  disk, reaching $n \sim 10^{8} \: {\rm cm^{-3}}$ after a few tens of AU. In simulations with minimum cell sizes worse than this, the limited spatial resolution will prevent the HII region from becoming trapped, regardless of whether or not it should do so in reality.

\subsection{Caveats}
Taken at face value, our results suggest that the HII regions produced by massive Pop.\ III stars will generally be trapped in close proximity to the stars, owing to the high gas densities surrounding them and the inability of the ionising radiation to clear this dense gas away. If true, this would imply that photoionisation feedback is not effective at shutting off accretion onto massive Pop.\ III stars, a result that would have important implications for predictions of the Pop.\ III initial mass function.

However, there are several reasons why one should treat this result with caution. First, although our simulations achieve sub-AU scale resolution of the gas flows in the vicinity of each massive Pop.\ III star, they do not achieve the same spatial resolution for the gravitational force exerted on the gas by the stars. For reasons of computational efficiency, the gravitational force from each sink particle in the simulation is softened on a scale $r_{\rm acc, soft} = r_{\rm sink} / 6$. Therefore, gas that is separated from the sink by less than $r_{\rm sink, soft}$ feels a weaker gravitational attraction to it than it would in reality. It is plausible that this leads to the accretion disk in the immediate vicinity of the star being thicker in our simulations than it would be in reality, thereby presenting more of an obstacle to the escape of the HII region. Comparison of the results from our simulations with $r_{\rm sink} = 10$~AU and $r_{\rm sink} = 30$~AU does not show any evidence that this is the case (see Appendix~\ref{app1}). However, a complete assessment of the extent to which the softening of the sink gravity affects the structure of the disk near the star and the interaction between the HII region and the disk will ultimately require simulations with much smaller values for $r_{\rm sink, soft}$ than those adopted here. Unfortunately, although such simulations are feasible when pursued for short periods -- see e.g.\ our own T2 and R2 runs -- their extremely high computational cost means that is it not currently possible to run them for long enough to form any stars massive enough to produce significant numbers of ionising photons. 

Second, our simulations are missing two important physical processes that may help to clear gas away from the massive stars, thereby aiding the escape of the HII region from the disk. The first of these is the magnetic field. Prior work has shown that even if one starts with a very weak seed magnetic field in a minihalo, produced e.g.\ by the Biermann battery \citep{biermann_uber_1950}, then the turbulent dynamo will quickly amplify the field to a strength at which it may become dynamically important \citep{kulsrud1997,schober_small-scale_2012,Liao2021,Sharda2021}. If this dynamo-amplified field is then wound up further in the rotating accretion disk, it may start to drive a bipolar outflow capable of removing at least some of the dense gas from the vicinity of the star \citep{tan2004,latif2016}. MHD simulations of Pop.\ III star formation carried out by \citet{machida_first_2006,machida_magnetohydrodynamics_2008} find that a bipolar outflow is always launched provided that the initial magnetic energy of the collapsing gas exceeds its initial rotational energy. However, these studies were carried out assuming an initially uniform magnetic field, oriented in the same direction as the angular momentum vector of the gas, and it remains to be seen whether the strongly tangled magnetic field produced by the turbulent dynamo is equally effective at driving an outflow.

The other important physical process missing from our current study is radiation pressure due to Lyman-$\alpha$ photon scattering. Owing to the very high optical depth of the accretion disk, Lyman-$\alpha$ photons that are produced by the massive star or by recombination in the HII region will scatter many times before escaping. The radiation pressure that they exert on the gas is therefore enhanced by a factor of $N_{\rm scat}$, the mean number of scatterings that a Lyman-$\alpha$ photon undergoes before it escapes \citep{bithell1990,oh2002}. In the limit of high optical depth, this is given by $N_{\rm scat} \sim 15 (\tau_{0} / 10^{5.5})^{1/3}$, where $\tau_{0}$ is the optical depth of the Lyman-$\alpha$ line at line centre \citep{adams1972,adams1975,milos2009}. If we assume that Lyman-$\alpha$ scattering within the HII region itself is unimportant \citep{stacy_building_2016}, then the dominant contribution to $\tau_{0}$ comes from the shell of atomic hydrogen surrounding the HII region. This is not resolved in our simulations, but a plausible upper limit on its column density is $N_{\rm H} \sim 10^{25} \: {\rm cm^{-2}}$, as much thicker shells are highly opaque to Lyman-Werner photons and hence are unlikely to remain fully atomic \citep{glover_dynamics_2017}. The Lyman-$\alpha$ optical depth can be written in terms of $N_{\rm H}$ as $\tau_{0} = \sigma_{0} N_{\rm H}$ where $\sigma_{0} = 5.9 \times 10^{-14} T_{4}^{-1/2} \: {\rm cm^{2}}$, with $T_{4} = T / 10^{4} \: {\rm K}$ \citep{milos2009}. Therefore, for an atomic hydrogen column density of $10^{25} \: {\rm cm^{-2}}$, we have $\tau_{0} \sim 2 \times 10^{12}$, and hence $N_{\rm scat} \sim 2700$. If we account for the fact that only two-thirds of recombinations produce a Lyman-$\alpha$ photon \citep{osterbrock2006} and that the energy of these photons is smaller than the mean energy of the ionising photons, then the end result is that the Lyman-$\alpha$ radiation pressure could plausibly be as much as a thousand times larger than the radiation pressure due to the ionising photons. Such a substantial enhancement in the radiation pressure force would bring it close in value to the gravitational attraction of the star (see Figure~\ref{fig:icsFigures-gravPotential}). This may make it sufficiently strong to overcome the trapping of the HII region in the disk, as suggested by \citet{mckee_formation_2008}, but establishing whether or not this actually occurs will require a far more careful treatment of the interaction between the radiation field, the gas and the gravitational potential than can be provided by this simple estimate.

In view of these caveats, it is clear that we cannot safely conclude that the HII regions produced by massive Pop.\ III stars will necessarily remain trapped close to these stars, and we do not argue that this is the case. Rather, our simulations starkly highlight two important facts. First, {\em we do not currently understand how the early evolution of the HII regions produced by massive Pop.\ III stars proceeds}. Do the regions escape from the dense protostellar accretion disk because the disk becomes very thin close to the star, or because magnetically-driven outflows and/or radiation pressure clear away the gas, leaving a path free for the radiation to escape? How quickly does this occur? Does it, indeed, occur at all, or do the HII regions instead remain trapped for an extended period? Answering these questions will require simulations with higher resolution and a smaller sink accretion radius than those we present here, and that moreover treat the effects of the magnetic field and the scattering of Lyman-$\alpha$ photons. Carrying out such simulations for a long enough period to allow at least a few massive stars to form -- typically, of the order of $10^{4} \: {\rm yr}$, given typical Pop.\ III accretion rates -- is an extremely demanding computational challenge but is vital for making further progress on this topic. Second, our results show that {\em current simulations do not correctly model HII region breakout}. The typical numerical approach used to inject radiation in simulations of feedback from massive Pop.\ III stars places it in regions that lie outside of the dense accretion disk in the polar directions, i.e.\ it implicitly assumes that break-out of the HII region from the disk has already occurred. Our results show that making this assumption leads to qualitatively different behavior from that which occurs if we inject radiation at the actual locations of the massive stars, suggesting that this approach should only be used with extreme caution. 

Finally, it is also interesting to compare our results to simulations of the formation of supermassive stars (SMSs) in H$_{2}$-free halos. These models typically also find that the HII regions produced by the SMSs are trapped close to the stars and provide little impediment to ongoing accretion \citep{regan2020,sakurai2020}.

\section{Conclusions}
\label{sec:conclusions}

In this paper, we have presented results from a set of high resolution simulations of the formation of massive Pop.\ III stars. Starting with idealized initial conditions -- a super-critical Bonnor-Ebert sphere with a velocity field comprised of a rotational component and a turbulent component -- we followed the gravitational collapse of the gas and the formation and fragmentation of a dense, flattened accretion disk. We used sink particles to represent stars forming in this disk and followed the evolution of the system for a period of 20~kyr after the formation of the first protostar. The simulation was repeated several times with different combinations of turbulent and rotational energy and different values for the sink accretion radius. For each combination of the initial  settings, we ran one simulation without radiative feedback and two with radiative feedback: a fiducial model in which the attenuation of radiation on scales less than the sink accretion radius was properly accounted for and a variant model in which this attenuation was artificially suppressed. The latter model was included in our study as it better represents the approach used in a number of previous studies of Pop.\ III star formation. The main results of our simulations can be summarised as follows:

\begin{itemize}
\item In our fiducial model, the ionising photons produced by the massive Pop.\ III stars are absorbed in the dense gas immediately surrounding the stars. The HII regions produced by these stars are very small ($< 1$~AU) and are trapped within the dense gas. This occurs despite the pressure gradient that exists between the ionised and the neutral gas because the pressure of the ionised gas is overwhelmed by the gravitational attraction of the star itself. Accounting for the radiation pressure of the ionising photons does not change this conclusion. 
\item Lyman-Werner photons produced by the massive stars can escape from the dense gas if the atomic hydrogen column density is low enough, but have little impact on the gas dynamics.
\item Because of the trapping of the HII region, the inclusion of radiative feedback has no significant impact on either the number or the total mass of protostars formed.
\item Artificially suppressing attenuation within the sink accretion radius allows ionising radiation to escape from the dense accretion disk surrounding the star, resulting in the growth of a bipolar HII region. This occurs because the size of the sink accretion radius or sink cell used in most calculations is larger than the local scale height of the disk.
\end{itemize}

Our results clearly demonstrate that the interaction between the ionising radiation produced by massive Pop.\ III stars and their birth environment is highly sensitive to the details of how the radiation is injected into the simulation. Injecting the radiation at too large a distance from the actual location of the star (as in our variant model) allows it to avoid the dense gas near the star and hence artificially prevents HII region trapping from occurring. Whether HII region trapping occurs in reality remains unclear. Confirming this will require simulations that have even better resolution of the gravitational potential near the massive stars (to ensure they recover the correct disk scale height) and that account for the effects of magnetic fields and Lyman-$\alpha$ radiation pressure. Until such simulations become possible, the question of how the HII regions produced by Pop.\ III stars evolve at early times will remain unresolved.

\section*{Acknowledgements}
The authors thank an anonymous referee for insightful comments, and they acknowledge stimulating discussions with Robin Tre{\ss}, Anna Schauer, Mattis Magg, Paul C.\ Clark, Philipp Girichidis, John Regan, Zoltan Haiman and Kazuyuki Sugimura on various technical and scientific aspects of this work. RSK and SCOG acknowledge funding from the European Research Council via the ERC Synergy Grant ``ECOGAL -- Understanding our Galactic ecosystem: From the disk of the Milky Way to the formation sites of stars and planets'' (project ID 855130), from the Deutsche Forschungsgemeinschaft (DFG) via the Collaborative Research Center (SFB 881, Project-ID 138713538) ``The Milky Way System'' (sub-projects A1, B1, B2 and B8) and from the Heidelberg cluster of excellence (EXC 2181 - 390900948) ``STRUCTURES: A unifying approach to emergent phenomena in the physical world, mathematics, and complex data'', funded by the German Excellence Strategy. SG acknowledges financial support from a NOVA grant for the theory of massive star formation. The project made use of computing resources provided by {\sl The L\"{a}nd} through bwHPC and by DFG through grant INST 35/1134-1 FUGG. Data are in part stored at SDS@hd supported by the Ministry of Science, Research and the Arts Baden-Württemberg (MWK) and by DFG through grant INST 35/1314-1 FUGG.

\section*{Data Availability}
The simulation data used within this paper will be shared on reasonable request to the corresponding author.




\bibliographystyle{mnras}
\bibliography{bib/thesis_zotero.bib,bib/other.bib}



\appendix

\section{Assessing the influence of the gravitational softening on the disk structure}
\label{app1}

\begin{figure}
  \includegraphics[width=\linewidth]{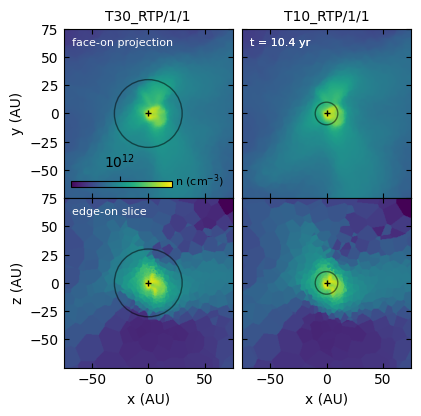}
  \caption{Face-on projection (top) and edge-on slices (bottom) of the gas number density around the first sink to form in simulations T30\_RTP (left) and T10\_RTP (right) at a time $t = 10.4$~yr after the formation of the sink. In the case of the projections we plot the column-averaged number density. The position of the sink is marked with a small black cross and the sink accretion radius is indicated with a black circle. We see from the plot that although the gravitational softening length adopted for the sink particle differs by a factor of three between the two runs, this does not significantly affect the distribution of the gas close to the sink.}
  \label{fig:sinksoft}
\end{figure}

To assess the extent to which the softening of the sink gravity might affect the density structure of the gas close to the sink accretion radius, we have compared the density structure around the first sink to form in the T10\_RTP and and T30\_RTP simulation. Prior to sink formation, the evolution of these simulations is identical, meaning that the first sink forms at the same location in both runs, allowing for a meaningful comparison between them. In Figure~\ref{fig:sinksoft}, we show face-on and edge-on views of the density structure around the sinks at a time $t = 10.4$~yr after sink formation.\footnote{Note that although this time is early in the evolution of the system, it nevertheless corresponds to approximately two dynamical times at the sink creation density, and so the gas will have had plenty of time to respond to the presence of the sink.} If the gravitational softening were to have a significant effect on the density structure close to the accretion  radius, then we should see clear differences between the structure on scales of 10 to 30 AU in the two simulations, since these scales are outside of the accretion radius in run T10\_RTP but within it in run T30\_RTP. In practice, we see that there is very little difference between the two runs on these scales at this early time, suggesting that the gravitational softening does not significantly affect the structure on these scales. We cannot rule out the softening having a greater effect on smaller scales ($r \ll 10$~AU), as here the structure of the gas will also be strongly affected by accretion onto the sink particle. In addition, we note that we cannot carry out a similar comparison for the sinks forming at later times in the simulations, as the later evolution of the accretion disk is nonlinear and does not proceed in exactly the same fashion in both simulations, preventing an simple 1:1 comparison.

\bsp	
\label{lastpage}
\end{document}